\documentclass[%
superscriptaddress,
reprint,
amsmath,amssymb,
aps,
pre,
]{revtex4-2}

\usepackage[english]{babel}

\usepackage{graphicx}% Include figure files
\usepackage[percent]{overpic}
\usepackage{dcolumn}% Align table columns on decimal point
\usepackage{bm}% bold math

\usepackage{hyperref}% add hypertext capabilities
\usepackage{color}
\definecolor{rltred}{rgb}{0.75,0,0}
\definecolor{rltgreen}{rgb}{0,0.6,0}
\definecolor{rltblue}{rgb}{0.3,0.3,1}

\usepackage{placeins}

\hypersetup{colorlinks,%
	hypertexnames=true,%
	linkcolor=rltred,%
	citecolor=rltgreen,%
	linktocpage=true}

\begin{document}
\author{Mahdi Kourehpaz}
\affiliation{Institute for Theoretical Physics, Vienna University of Technology,
	Wiedner Hauptstra\ss e 8-10/136, 1040 Vienna, Austria, EU}

\author{Stefan Donsa}
\affiliation{Institute for Theoretical Physics, Vienna University of Technology,
	Wiedner Hauptstra\ss e 8-10/136, 1040 Vienna, Austria, EU}

\author{Fabian Lackner}
\affiliation{Institute for Theoretical Physics, Vienna University of Technology,
	Wiedner Hauptstra\ss e 8-10/136, 1040 Vienna, Austria, EU}

\author{Joachim Burgd\"orfer}
\affiliation{Institute for Theoretical Physics, Vienna University of Technology,
	Wiedner Hauptstra\ss e 8-10/136, 1040 Vienna, Austria, EU}

\author{Iva B\v rezinov\'a}
\email{iva.brezinova@tuwien.ac.at}
\affiliation{Institute for Theoretical Physics, Vienna University of Technology,
	Wiedner Hauptstra\ss e 8-10/136, 1040 Vienna, Austria, EU}

\title{Canonical density matrices from eigenstates of mixed systems}

\begin{abstract}
One key issue of the foundation of statistical mechanics is the emergence of equilibrium ensembles in isolated and closed quantum systems. Recently, it was predicted that in the thermodynamic ($N\rightarrow\infty$) limit of large quantum many-body systems canonical density matrices emerge for small subsystems from almost all pure states. This notion of canonical typicality is assumed to originate from the entanglement between subsystem and environment and the resulting intrinsic quantum complexity of the many-body state. For individual eigenstates it has been shown that local observables show thermal properties provided the eigenstate thermalization hypothesis holds, which requires the system to be quantum chaotic. In the present paper, we study the emergence of thermal states in the regime of a quantum analog of a mixed phase space. Specifically, we study the emergence of the canonical density matrix of an impurity upon reduction from isolated energy eigenstates of a large but finite quantum system the impurity is embedded in. Our system can be tuned by means of a single parameter from quantum integrability to quantum chaos and corresponds in between to a system with mixed quantum phase space. We show that the probability for finding a canonical density matrix when reducing the ensemble of energy eigenstates of the finite many-body system can be quantitatively controlled and tuned by the degree of quantum chaos present. For the transition from quantum integrability to quantum chaos we find a continuous and universal (i.e.~size independent) relation between the fraction of canonical eigenstates and the degree of chaoticity as measured by the Brody parameter or the Shannon entropy.
\end{abstract}
%%%%%%%%%%%%%%%%
\maketitle

%%%%%%%%%%%%%%%%%%%%%%%%%%%%%%%%%%%%%%%%%%%%%%%%%%%%%%%%%%%%%%%%%%%%%%%%%%
\section{Introduction}\label{sec:intro}
%%%%%%%%%%%%%%%%%%%%%%%%%%%%%%%%%%%%%%%%%%%%%%%%%%%%%%%%%%%%%%%%%%%%%%%%%%
As first recognized by Ludwig Boltzmann \cite{boltzmann_1896, huang_1988} ``molecular" chaos lies at the core of the foundation of classical statistical mechanics. Only when the phase space of an isolated mechanical system is structureless can the motion be safely assumed to be ergodic and the equal a priori probability for phase space points on the energy hypersurface, the basic tenet of the microcanonical ensemble, is realized. Moreover, chaotic dynamics is ``mixing", thereby enforcing the approach to the thermal equilibrium state from ``almost all" out-of-equilibrium initial conditions. While any large isolated system is expected to be described by a microcanonical ensemble, any well-defined small subsystem thereof that is only allowed to exchange energy with the remainder of the large system {described by the microcanonical ensemble (referred to as bath or environment in the following) is described by the canonical ensemble. The phase-space density of the subsystem is weighted by the Boltzmann factor $e^{-\beta H_s}$ where $H_s$ is the Hamilton function of the subsystem, $\beta=1/k_{\rm B}T$ with $T$ the temperature imprinted by the environment and $k_{\rm B}$ the Boltzmann constant. However, when the phase space of the system is not chaotic but rather dominated by regular motion on KAM tori \cite{arnold_1979, lichtenberg_1991}, neither ergodicity nor mixing is a priori assured and thermalization of an initial non-equilibrium state may be elusive. The implicit assumption of classical equilibrium statistical mechanics is that in the limit of a large number of degrees of freedom chaos is generic for any interacting many-particle system.

How those concepts translate into quantum physics has remained a topic of great conceptual interest and lively debate \cite{deutsch_quantum_1991,srednicki_chaos_1994, srednicki_thermal_1996, srednicki_approach_1999, zelevinsky_nuclear_1996, flambaum_towards_1996, flambaum_statistical_1997, borgonovi_chaos_1998, borgonovi_semiquantal_2002, borgonovi_quantum_2016, muller_thermalization_2015, reimann_why_2020, brenes_multipartite_2020, brenes_out--time-order_2021, sugimoto_eigenstate_2022, wang_complexity_2020, wang_quantum_2021}. Renewed interest is stimulated by the experimental accessibility of ultracold quantum gases \cite{braun_negative_2013,kaufman_quantum_2016,jansen_eigenstate_2019, abanin_colloquium_2019, prufer_observation_2018, erne_universal_2018,scherg_observing_2021}, trapped ions \cite{neyenhuis_observation_2017}, and nano-systems \cite{neill_ergodic_2016, trotzky_probing_2012} where many of the underlying concepts became quantitatively accessible in large but finite quantum systems in unprecedented detail. The foundation of thermalization of quantum systems has been pioneered by von Neumann in term of the quantum ergodic theorem \cite{neumann_beweis_1929, goldstein_long-time_2010, dalessio_quantum_2016, siskens_ergodicity_1973, bongaarts_observables_1974, mori_thermalization_2018}. Accordingly, the entropy is an increasing function of time and expectation values of generic macroscopic observables for pure states formed by coherent superposition of states within microscopic energy shells converge to that of the microcanonical ensemble provided that the energy spectrum of the system is strictly non-degenerate. Recently, this description of thermal equilibrium states was extended to the notion of canonical typicality \cite{popescu_entanglement_2006, goldstein_canonical_2006, reimann_typicality_2007, mori_thermalization_2018}. Accordingly, starting from almost any pure state formed by a coherent superposition of energy eigenstates of a large isolated many-body system with eigenenergies within a given energy shell $[E,E+\Delta E]$ of macroscopically small thickness $\Delta E$, the reduction to a small subsystem by tracing out the degrees of freedom of the bath will yield the same reduced density matrix one would obtain from the reduction of the microcanonical density matrix for the entire system. If the bath is sufficiently large and the interactions between the bath and the subsystem sufficiently weak, the reduced density matrix corresponds to the standard canonical density matrix $\hat \rho_s = e^{-\beta \hat H_s}/{\rm Tr}[ e^{-\beta \hat H_s}]$ with $\hat H_s$ the Hamilton operator of the subsystem. The proof of this canonical typicality invokes the intrinsic randomness of the expansion coefficients of the pure state in terms of entangled subsystem-bath states. The latter assumption goes back to the notion of intrinsic quantum complexity of entangled states in large systems put forward already by Schr\"odinger \cite{schroedinger_1965}.\\
An alternative approach to thermalization is tied to the eigenstate thermalization hypothesis (ETH) \cite{deutsch_quantum_1991, srednicki_chaos_1994, srednicki_thermal_1996, srednicki_approach_1999} first put forward by Landau and Lifshitz \cite{landau_1958} stating that basic properties of statistical mechanics can emerge not only from ensemble averages but from typical single wavefunctions. However, the condition under which such an equivalence may emerge has remained open. The more recent formulation of the ETH \cite{deutsch_quantum_1991, srednicki_chaos_1994, srednicki_thermal_1996, srednicki_approach_1999} invokes the notion of quantum chaos and Berry's conjecture. Characteristics of quantum chaos were originally identified in few-degrees of freedom systems whose classical limit exhibits chaos \cite{bohigas_characterization_1984, casati_stochastic_1979, casati_relevance_1987,haake_2001, gutzwiller_1990, berry_quantum_1989, prosen_energy_1993}. Nowadays, the notion of quantum chaos is invoked more generally for systems that display the same signatures such as energy level distributions predicted by random matrix theory (RMT) \cite{bohigas_characterization_1984, berry_level_1977, mehta_2004, berry_quantum_1989} or randomness of wavefunction amplitudes \cite{shapiro_onset_1984, deutsch_quantum_1991} even when a well-defined classically chaotic counterpart is not known. The ETH conjectures that for chaotic systems the diagonal matrix elements of any generic local observable taken in the energy eigenstate basis are smooth functions of the total energy while the off-diagonal elements are exponentially decreasing randomly fluctuating variables with zero mean \cite{srednicki_chaos_1994, srednicki_thermal_1996, srednicki_approach_1999}. If the ETH is valid for a specific system individual eigenstates show thermal properties upon reduction to a small subsystem. The ETH has been shown to hold for a large variety of systems without a classical analogue \cite{rigol_thermalization_2008, rigol_breakdown_2009, rigol_quantum_2009, rigol_alternatives_2012, santos_localization_2010, santos_onset_2010, genway_dynamics_2010, genway_thermalization_2012, kim_testing_2014, garrison_does_2018, schlagheck_dynamical_2016, jansen_eigenstate_2019, brenes_eigenstate_2020}. Deviations from the ETH have been observed for local observables in finite systems of hard-core bosons and spin-less fermions \cite{santos_localization_2010, santos_onset_2010, santos_weak_2012} when the energy level distribution deviates from the Wigner-Dyson level statistics of RMT characteristic for chaotic systems.\\
In the present paper we explore the quantitative relationship between thermal properties of reduced density matrices (RDMs) emerging from single isolated eigenstates of the entire system and quantum chaos. More specifically, we want to address the question: Is for large but finite systems quantum chaos a conditio sine qua non for the emergence of the Gibbs ensemble, i.e.~the canonical ensemble of the subsystem, from eigenstates of the entire system? Or is quantum entanglement and complexity in these systems itself sufficient to render the reduced density matrix of a small subsystem canonical? To this end, we determine the fraction of canonical density matrices emerging upon reduction from the entire set of eigenstates. We explore the existence of a quantitative relationship between the fraction of eigenstates that upon reduction lead to canonical eigenstates, also termed fraction of canonical eigenstates, and the degree of quantum chaos of the entire system. We unravel the connection between this eigenstate canonicity and quantum chaos by exact diagonalization of a large yet finite mesoscopic quantum system. We emphasize that this measure addresses isolated energy eigenstates of the many-body system, in contrast to coherent superpositions of energy eigenstates from a given energy shell of finite width with random expansion coefficients as invoked in the well-established notion of canonical typicality \cite{popescu_entanglement_2006, goldstein_canonical_2006, reimann_typicality_2007, mori_thermalization_2018}.\\
As a prototypical case in point we consider an itinerant impurity embedded in a spin-polarized Fermi-Hubbard system. Unlike impurity models for disordered systems \cite{krause_nucleation_2021} our model is fully deterministic. All key ingredients for the realization of the present system, i.e.~discrete lattice, tunable interactions, and impurity can be experimentally realized with ultracold fermionic atoms (see e.g. \cite{gaunt_bose-einstein_2013,fukuhara_quantum_2013,haller_single-atom_2015,parsons_site-resolved_2015,cheuk_quantum-gas_2015,greif_site-resolved_2016}). In the present scenario the impurity serves as a probe or ``thermometer" in the isolated many-body quantum system providing an unambiguous subsystem-bath decomposition with tunable coupling strength between subsystem and bath. Moreover, this system features a tunable transition from quantum chaos to quantum integrability without invoking any extrinsic stochasticity or disorder \cite{krause_nucleation_2021}. The fact that the subsystem consists of a distinguishable particle has a number of distinct advantages: The reduced density matrix of the probe is uniquely defined and its properties are basis independent. No choice of a specific basis for the probe such as the independent-particle basis is involved. Moreover, its canonical RDM approaches a Maxwell-Boltzmann rather than a Fermi-Dirac distribution for indistinguishable fermions. Its thermal state is thus characterized by a single equilibrium parameter, the temperature $T$, without the need for introducing a chemical potential $\mu$ thereby improving the numerical accuracy of the test of canonicity. We measure the proximity of the reduced density matrix of the impurity to the canonical density matrix and identify a direct and size-independent correlation between the fraction of canonical eigenstates and quantum chaos.\\
The paper is structured as follows. In Sec.~\ref{sec:qu_sys} we introduce our impurity-Fermi-Hubbard model which serves as a prototypical (sub)system-environment model system. Quantitative measures for quantum chaos are introduced in Sec.~\ref{sec:qmchaos}. The mapping of spectral properties of this isolated many-body system onto thermal states of the impurity within the framework of the microcanonical and canonical ensembles are discussed in Sec.~\ref{sec:1rdm}. The distance in Liouville space between the reduced density matrix of the impurity and a generic canonical density matrix is analyzed and the relation between the fraction of canonical eigenstates and quantum chaos is established in Sec.~\ref{sec:can_typ}. Concluding remarks are given in Sec.~\ref{sec:conc}.
%
%%%%%%%%%%%%%%%%%%%%%%%%%%%%%%%%%%%%%%%%%%%%%%%%%%%%%%%%%%%%%%%%%%%%%%%%%%
\section{The Fermi-Hubbard model with impurity}\label{sec:qu_sys}
%%%%%%%%%%%%%%%%%%%%%%%%%%%%%%%%%%%%%%%%%%%%%%%%%%%%%%%%%%%%%%%%%%%%%%%%%%
We investigate a variant of the single-band one-dimensional Fermi-Hubbard model which is particularly well suited to study entanglement and quantum correlations between subsystem and its environment or bath. The bath is represented by spin-polarized fermions enforcing single occupancy of sites by bath particles while the distinguishable impurity can occupy any site. Accordingly, the Hamiltonian of the total system is given by 
\begin{equation}\label{eq:total_ham}
\hat{H} = \hat{H}_{\rm I}+\hat{H}_{\rm B}+\hat{H}_{\rm IB},
\end{equation}
where the Hamiltonian of the subsystem, i.e.~the impurity (I), is 
\begin{eqnarray}\label{eq:imp_ham}
	\hat{H}_{\rm I}  &= -J_{\rm I} \sum_{j=1}^{M_s-1} \left[ \hat{a}^{\dagger}_{j+1}\hat{a}_{j} + c.c.\right] +  \sum_{j=1}^{M_s} V(j)\hat{n}_{j},
\end{eqnarray}
while the Hamiltonian of the bath is 
\begin{align}\label{eq:bath_ham}
\hat{H}_{\rm B}  = &-J_{\rm B} \sum_{j=1}^{M_s-1} \left[ \hat{b}^{\dagger}_{j+1}\hat{b}_{j} + c.c.\right] 
+ W_{\rm BB}\sum_{j=1}^{M_s-1} \hat{N}_{j+1}\hat{N}_{j} \nonumber \\
&+ \sum_{j=1}^{M_s} V(j)\hat{N}_{j}.
\end{align}
The interaction between the subsystem and the bath is given by
\begin{equation}\label{eq:interact}
\hat{H}_{\rm IB} = W_{\rm IB}\sum_{j=1}^{M_s}\hat{n}_j\hat{N}_j.
\end{equation}
The operators $\hat a_j$ and $\hat a_j^{\dagger}$ ($\hat b_j$ and $\hat b_j^{\dagger}$) are the creation and annihilation operators of the impurity (bath particles) on site $j$ with the anticommutation relations $\{a_i,a_j\}=0$, $\{a_i^\dagger,a_j^\dagger\}=0$, $\{a_i,a_j^\dagger\}=\delta_{ij}$, and $\{a_i^{(\dagger)},b^{(\dagger)}\}=0$. The operators $\hat n_j=\hat a_j^\dagger\hat a_j$ and $\hat N_j=\hat b_j^\dagger \hat b_j$ correspond to the number operators of impurity and bath $\hat n_j|j\rangle = n_j|j\rangle$ and $\hat N_j|j\rangle = N_j|j\rangle$ with occupation numbers $n_j$ and $N_j$ of site $j$, respectively. $J_{\rm I}$ ($J_{\rm B}$) describes the hopping matrix elements of the impurity (bath particles). The bath particles interact with each other by a nearest-neighbor interaction with strength $W_{\rm BB}$ while the impurity interacts with the bath particles via an on-site interaction with strength $W_{\rm IB}$. The Hubbard chain has $M_s$ sites with Dirichlet boundary conditions imposed at the edges. An additional very weak external background potential ($V\ll J_{\rm I}, J_{\rm B}$) with on-site matrix element $V(j)$ ($j=1,\ldots,M_s$) is applied, 
\begin{equation}\label{eq:pot}
	V(j)= 0.01\left[-0.5+\frac{(j-1)^n}{(M_s-1)^n}\right],
\end{equation}
for which we use a linear ($n=1$) or quadratic ($n=2$) function in order to remove residual geometric symmetries such that the irreducible state space coincides with the entire state space and symmetry related degeneracies are lifted. Alternative impurity models were recently suggested for the investigation of the ETH \cite{brenes_eigenstate_2020}.\\
We solve the system via exact diagonalization to determine all eigenstates and eigenenergies of the entire system. The dimension of the Hilbert space of the system is $d_{\rm H}=M_s{{M_s}\choose{N_{\rm B}}}$, where $N_{\rm B}$ is the number of bath particles. We consider typical half-filling configurations with $N_{\rm B} \approx M_s/2$. The largest $M_s$ considered is $M_s=15$ resulting in a Hilbert space dimension of 
$d_{\rm H}=96525$ for $N_{\rm B}=7$. We set $J_{\rm I}=J_{\rm B}=J$ which also defines the unit of energy ($J=1$) in the following.  The key advantage of the present model is that it allows to control and tune the properties of the bath separately by varying $W_{\rm BB}$ while keeping fixed the properties of the subsystem whose reduced density matrix we probe. This clear-cut subsystem-bath decomposition allows for the unambiguous probing of the emergence of canonical density matrices thereby avoiding any ad-hoc separation by ``cutting out" of the subsystem which then requires the grand canonical density matrix for an open quantum system since both energy and particles can be exchanged\cite{santos_weak_2012}. Moreover, its thermal state is unambiguously characterized by $T$ rather than by $T$ and $\mu$ as for indistinguishable fermions thereby improving the numerical reliability of the performed tests.\\
The present system should be realizable for ultracold fermionic atoms trapped in optical lattices \cite{parsons_site-resolved_2015,haller_single-atom_2015,cheuk_quantum-gas_2015,schreiber_observation_2015,boll_spin-_2016, greif_site-resolved_2016,scherg_observing_2021}. All key ingredients required for its realization including tunable interactions and impurity-bath mixtures are available in the toolbox of ultracold atomic physics. We note that tuning the nearest-neighbor interaction $W_{\rm BB}$ between the atoms in optical lattices to large values in the regime of strong correlations, $W_\text{BB}/J_\text{B}\gtrsim 1$, still poses an experimental challenge which might be overcome in the near future.
%%%%%%%%%%%%%%%%%%%%%%%%%%%%%%%%%
\section{Measures of quantum chaos}\label{sec:qmchaos}
%%%%%%%%%%%%%%%%%%%%%%%%%%%%%%%%%
The present single-band Fermi-Hubbard model does not possess an obvious classical counterpart whose phase space consists of regions of regular and/or chaotic motion. Lacking such direct quantum-classical correspondence, quantum integrability and quantum chaos in the present system is identified by signatures of the quantum system that have been shown to probe chaotic and regular motion in systems where quantum-classical correspondence does prevail. Several measures of quantum chaos have been proposed that are based on either properties of eigenstates or of the spectrum \cite{berry_regular_1977, prosen_energy_1993, borgonovi_quantum_2016, pandey_adiabatic_2020, santos_onset_2010, wang_statistical_2020, lozej_quantum_2022, li_statistical_1994}. As will be shown below, by tuning $W_{\rm BB}$ we can continuously tune the entire system from the limit of quantum integrability to the limit of quantum chaos across the transition region of a mixed quantum system in which integrable and chaotic motion coexist and explore its impact on the fraction of eigenstates which upon reduction lead to canonical density matrices. The influence of the continuous transition from quantum integrability to quantum chaos on the thermal state of the subsystem will be explored with the help of the present prototypical system.
%%%%%%%%%%%%%%%%%%%%%%%%%%%%%%%%%%%
\subsection{Spectral measures}\label{subsec:spec}
%%%%%%%%%%%%%%%%%%%%%%%%%%%%%%%%%%%
Starting point for analyzing and quantifying quantum chaos by means of spectral statistics is the cumulative spectral function also called the staircase function 
\begin{equation}\label{eq:stcf}
	N(E) = \sum_\alpha \Theta(E-E_\alpha),
\end{equation}
where $E_\alpha$ are the energy eigenvalues of the entire system, and $\Theta$ is the Heaviside step function. Its spectral derivative is the density of states (DOS)
\begin{equation}\label{eq:dos}
	\Omega(E)=\frac{d}{dE}N(E).
\end{equation}
Examples for $N(E)$ and $\Omega(E)$ of the present system are shown in Fig.~\ref{fig:dos}.
\begin{figure}[t]
	\includegraphics[width = 8 cm]{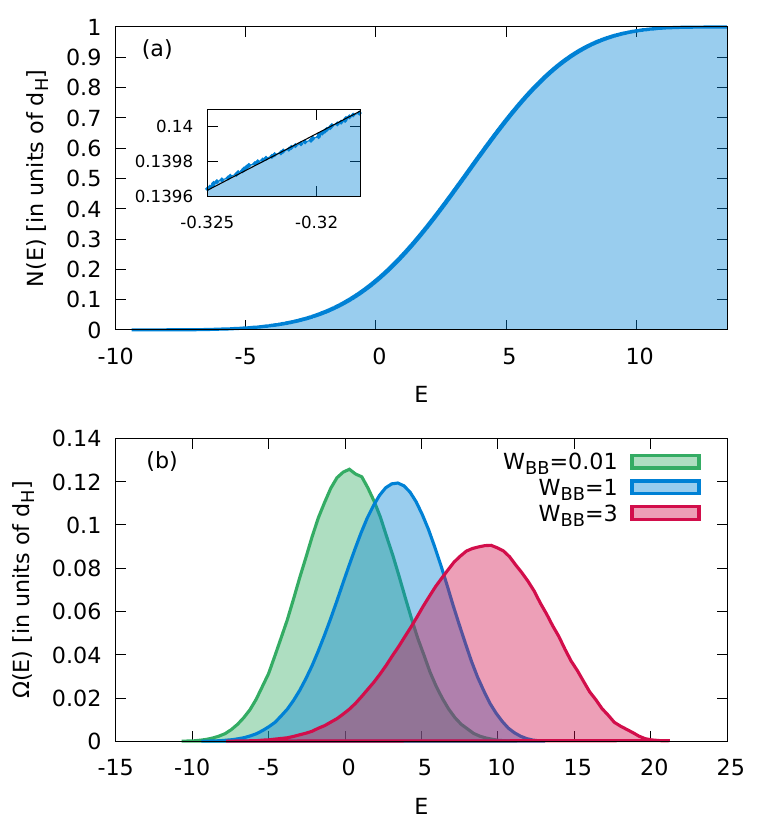}
	\caption{(a) The spectral staircase function $N(E)$ for the Fermi-Hubbard model with impurity with $W_{\rm BB}=1$ and a total number of states $d_{\rm H}=96525$ ($M_s=15$, $N_{\rm B}=7$). The inset shows a magnification of $N(E)$ with a fit for the smoothed ``average" staircase function $\bar N(E)$ entering the spectral unfolding. (b) The normalized DOS, $\Omega(E)$, using a bin size of $\Delta E=0.4$ for different interaction strengths of the bath particles. The impurity-bath interaction in (a) and (b) is $W_{\rm IB}=1$.}
	\label{fig:dos}
\end{figure}
The smoothed ``average" spectral staircase function $\bar N(E)$ fitted to a polynomial of order $10$, also shown in Fig.~\ref{fig:dos}, provides the reference for spectral unfolding required for certain measures of quantum fluctuations about the (classical) mean. Accordingly, the unfolded energy spectrum is given by $e_\alpha = \bar N(E_\alpha)$. For systems for which quantum-classical correspondence holds, $\bar N(E)$ corresponds to the classical phase space volume in units of Planck's constant $h$ and $\Omega(E)$ to the microcanonical energy shell. We note that the saturation of $N(E)$ observed with increasing $E$ [Fig.~\ref{fig:dos} (a)] or, likewise, the bell-shaped curve for the DOS [Fig.~\ref{fig:dos} (b)] decreasing at large $E$ is in the present case a consequence of the single-band approximation of the Fermi-Hubbard model (Eq.~\ref{eq:total_ham}) and, more generally, appears for systems with a spectrum bounded from above. For realistic macroscopic systems, $N(E)$ and $\Omega(E)$ should, generically, increase monotonically with $E$. As discussed in more detail below, this non-generic decrease of the density of states observed for the present as well as for other finite and mesoscopic systems has implications for the ensuing thermal properties.  \\
The probability density $P(s)$ of the nearest-neighbor level spacings (NNLS), $s=e_{\alpha+1}-e_\alpha$, features distinctively different shapes for quantum integrable and quantum chaotic systems. While for integrable systems the NNLS have been predicted by Berry and Tabor \cite{berry_level_1977} to feature an exponential (or Poissonian) distribution $P_{\rm P}(s) = \exp{(-s)}$, for chaotic systems it closely follows random matrix theory \cite{bohigas_characterization_1984}. In our case of a time-reversal symmetric system the corresponding random-matrix ensemble is the Gaussian orthogonal ensemble (GOE) which has been shown (see e.g.~\cite{haake_2001}) to closely follow the Wigner-Dyson distribution (or Wigner surmise) given by
\begin{equation}\label{eq:wigner_dyson}
	P_{\rm WD}(s) = \frac{\pi s}{2}e^{-\pi s^2/4}.
\end{equation}
A complementary spectral measure first proposed by Gurevich and Pevzner \cite{gurevich_57} and applied to quantum chaos \cite{oganesyan_localization_2007, atas_distribution_2013} has the advantage that it does not require spectral unfolding but can be applied to the spectral raw data, i.e.~the restricted gap ratios $r_\alpha$
\begin{equation}\label{eq:r_alpha}
	r_\alpha = {\rm min}\left(r_\alpha, \frac{1}{r_\alpha}\right),
\end{equation}
where $r_\alpha=(E_{\alpha+1}-E_\alpha)/(E_\alpha-E_{\alpha-1})$. The distribution of restricted gap ratios has been shown to obey for $3\times3$ GOE matrices the analytical prediction
\begin{equation}\label{eq:w_goe}
	W_{\rm GOE}(r) =  \frac{27}{4}\frac{r+r^2}{(1+r+r^2)^{5/2}}.
\end{equation}
For chaotic systems this prediction remains very accurate even for large systems \cite{atas_distribution_2013}. In the limit of quantum integrable systems, the distribution of restricted gap ratios is given by (see \cite{oganesyan_localization_2007})
\begin{equation}\label{eq:w_p}
	W_{\rm P}(r) = \frac{2}{(1+r)^2}.
\end{equation}
The search for generic spectral measures for the transition regime between the quantum integrable and quantum chaotic limit has remained an open problem. For systems possessing a classical counterpart with a mixed phase space in which integrable and chaotic motion coexist, several models for the NNLS have been proposed \cite{berry_semiclassical_1984, hasegawa_stochastic_1988, caurier_level_1990, lenz_reliability_1991, izrailev_quantum_1988}. Empirically, one of the best fits to spectral data for mixed systems has been provided by a heuristic ansatz suggested by Brody \cite{brody_random-matrix_1981} which allows for a one-parameter smooth interpolation of the NNLS distribution in the transition region between the quantum integrable and quantum chaotic limit,
\begin{eqnarray}\label{eq:brody}
	P_{\rm B}(s) = (\gamma+1)bs^\gamma e^{-bs^{\gamma +1}}
\end{eqnarray}
where the Brody parameter $\gamma$ characterizes the transition from the integrable ($\gamma=0$) to the chaotic limit ($\gamma=1$) and $b$ follows from the normalization as
\begin{eqnarray}
	b = \left[\Gamma\left(\frac{\gamma+2}{\gamma+1}\right)\right]^{\gamma +1}.
\end{eqnarray}
The Brody parameter can be viewed as a measure of the strength of level repulsion between neighboring levels of the quantum system. For mixed few-degrees of freedom systems with a classical analogue, $\gamma$ has been found to be, also, a measure for the chaotic fraction of classical phase space \cite{lenz_reliability_1991, yang_molecular-dynamics_1991}. Moreover, $\gamma$ has also been found to be directly proportional to the degree of phase-space (de)localization of eigenstates as measured by their Husimi distribution \cite{wang_statistical_2020}. The parametrization of the transition from quantum integrability to quantum chaos in terms of a variable exponent $\gamma$ has the salient feature that even for very small but finite $\gamma$, $0<\gamma\ll1$, $P_{\rm B}(0)=0$, reflecting the fact that any perturbation of quantum integrability immediately causes level repulsion and suppresses the probability density for any exact degeneracy. We recall that non-degeneracy is one of the key prerequisites of von Neumann's quantum ergodic theorem \cite{neumann_beweis_1929}. We further note that the Hasegawa distribution \cite{hasegawa_stochastic_1988} sometimes provides an even more accurate fit to the NNLS distribution (see, e.g. \cite{libisch_graphene_2009}), however, at the price of a second adjustable parameter.\\
To determine $\gamma$ we fit Eq.~\ref{eq:brody} to the data for $P(s)$ (Fig.~\ref{fig:level_stat}). The quality of the fit is evaluated through the $\chi^2$-function 
\begin{equation}\label{eq:chi2}
	\chi^2 = \sum_i [P(s_i)-P_{\rm B}(s_i)]^2 \Delta s
\end{equation}
which measures the deviation of the distribution of nearest-neighbor spacings $P(s)$ from the Brody distribution $P_\text{B}(s)$ using a bin size of $\Delta s$. As an additional measure for the uncertainty of $\gamma$ we use the fact that the Brody parameter can be alternatively determined from a fit to the integral $\int ds'P(s')$ rather than to $P(s)$ itself. The small differences found between the two fits can be used as a measure for the numerical error.\\
\begin{figure}[t]
	\includegraphics[width=8 cm]{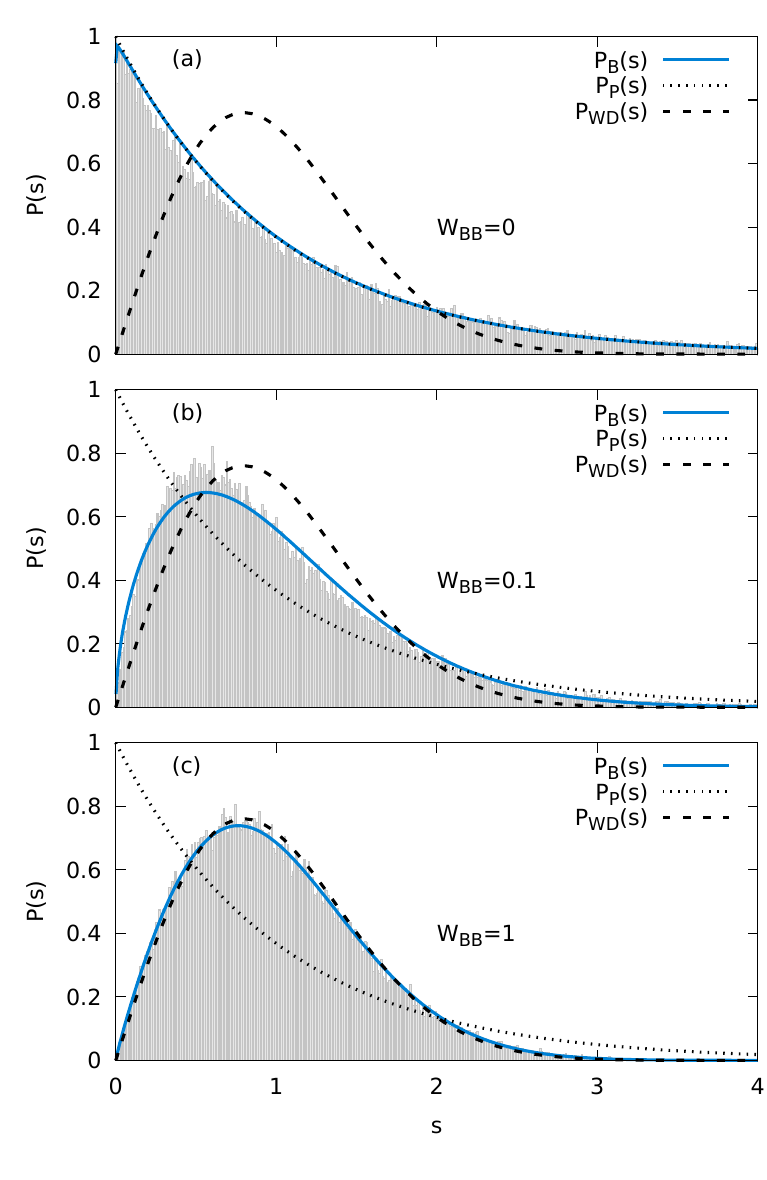}
	\caption{The numerically determined nearest-neighbor level statistics $P(s)$ for the Fermni-Hubbard model with impurity (Eqs.~\ref{eq:total_ham}-\ref{eq:interact}) (a) $W_{\rm BB}=0$, (b) $W_{\rm BB}=0.1$ and (c) $W_{\rm BB}=1$ compared to the Poisson (exponential) distribution $P_{\rm P}(s)$, the Wigner-Dyson distribution $P_{\rm WD}(s)$, as well as the fit to the Brody distribution $P_{\rm B}(s)$. The bin size used is $\Delta s =0.01$. Other parameters are $M_s=15$, $N_{\rm B}=7$, and $W_{\rm IB}=1$.}
	\label{fig:level_stat}
\end{figure}
For $W_{\rm BB}=0$ and for the linear tilt of the external potential $V(j)$ (Eq.~\ref{eq:pot}) we observe an excess of (near) degenerate states as compared to the prediction of the exponential (Poisson) distribution in the first bin at $s=0$ with $\Delta s=0.01$. This hints at the presence of an only weakly broken symmetry which disappears when using a quadratic tilt. For reasons of consistency, we employ for all $W_{\rm BB}$ a linear tilt in the following. Neglecting the first bin in the fitting procedure for $W_{\rm BB}=0$ we obtain $\gamma=0.005$ and, overall, a very good agreement to the Poisson distribution [Fig.~\ref{fig:level_stat} (a)].  As the intra-bath interaction is varied from $W_{\rm BB}=0$ to $W_{\rm BB}=1$ we observe a continuous transition from a near Poissonian to an approximate Wigner-Dyson NNLS distribution [Fig.~\ref{fig:level_stat} (a)-(c)]. The Brody parameter monotonically increases from $\gamma\simeq0.005$ at $W_{\rm BB}=0$ to $\gamma\simeq0.9$ at $W_{\rm BB}=1$. We note that after reaching a plateau at $\gamma\simeq0.93$ near $W_{\rm BB}=3$, the Brody parameter decreases again for $W_{\rm BB}> 5$ and vanishes in the strongly correlated limit of $W_{\rm BB}\gg1$. The decrease of the Brody parameter for large $W_{\rm BB}$ results from clustering of the energy spectrum in the strongly interacting regime. The bath fragments into clusters of particles with the interactions between separate clusters suppressed. Thus, a partially ordered system emerges reducing the degree of quantum chaoticity. We will focus in the following on the parameter range $W_{\rm BB}\leq1$ within which the transition from a nearly quantum integrable to a nearly fully quantum chaotic system occurs.\\
For the two limiting cases of quantum integrability ($W_\text{BB}\rightarrow 0$) and quantum chaos ($W_\text{BB}\rightarrow 1$) of the present Fermi-Hubbard system we can also apply the predictions for the restricted gap ratio distribution (Eq.~\ref{eq:w_goe}, Eq.~\ref{eq:w_p}). We find for these two limiting cases very good agreement between the prediction and the data (Fig.~\ref{fig:W}) confirming that the identification of quantum integrability and quantum chaos is independent of the particular choice of the spectral measure.
\begin{figure}[t]
	\includegraphics[width=\columnwidth]{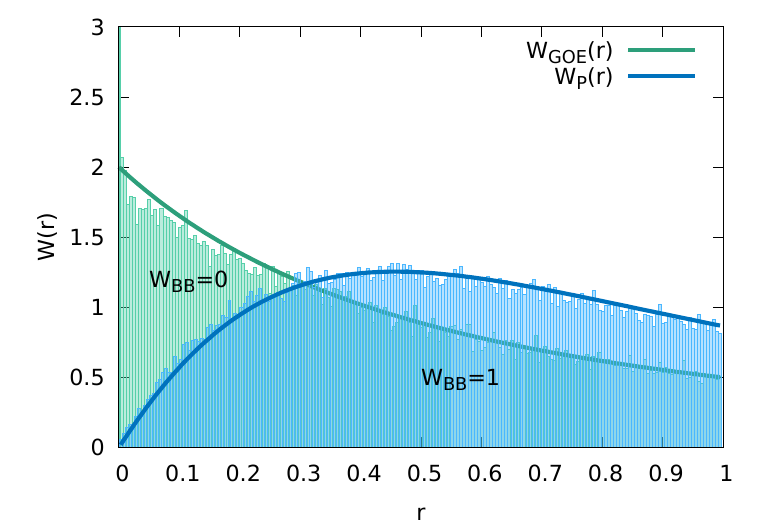}
	\caption{Statistical distribution function of restricted gap ratios for the Fermi-Hubbard model with impurity ($M_s=15$, $N_{\rm B}=7$) for different $W_{\rm BB}$ and $W_{\rm IB}=1$ compared to the analytical predictions for random matrices within the GOE ensemble (Eq.~\ref{eq:w_goe}) and for integrable spectra (Eq.~\ref{eq:w_p}).}
	\label{fig:W}
\end{figure}
For the first moment of the restricted gap ratio distribution we find $\langle r \rangle = 0.5284$ for $W_{\rm BB}=1$ agreeing to within $0.5\%$ with the GOE expectation value for asymptotically large matrices $\langle r \rangle_{\rm GOE}=0.5307$
\cite{atas_distribution_2013}. Conversely, for $W_{\rm BB} =0$ we find $\langle r \rangle = 0.3811$ in very good agreement with the prediction for a Poisson distribution $\langle r \rangle_{\rm P} = 0.3863$. As there is presently no interpolation function $W(r)$ available for the transition between the quantum integrable limit (Eq.~\ref{eq:w_p}) and the quantum chaotic limit (Eq.~\ref{eq:w_goe}), we will focus in the following on the Brody distribution for the NNLS as spectral measure for the transition regime.
%%%%%%%%%%%%%%%%%%%%%%%%%%%%%%%%%
\subsection{Measures for wavefunctions}\label{subsec:wfn}
%%%%%%%%%%%%%%%%%%%%%%%%%%%%%%%%%
%
\begin{figure}[t]
	\includegraphics[width=\columnwidth]{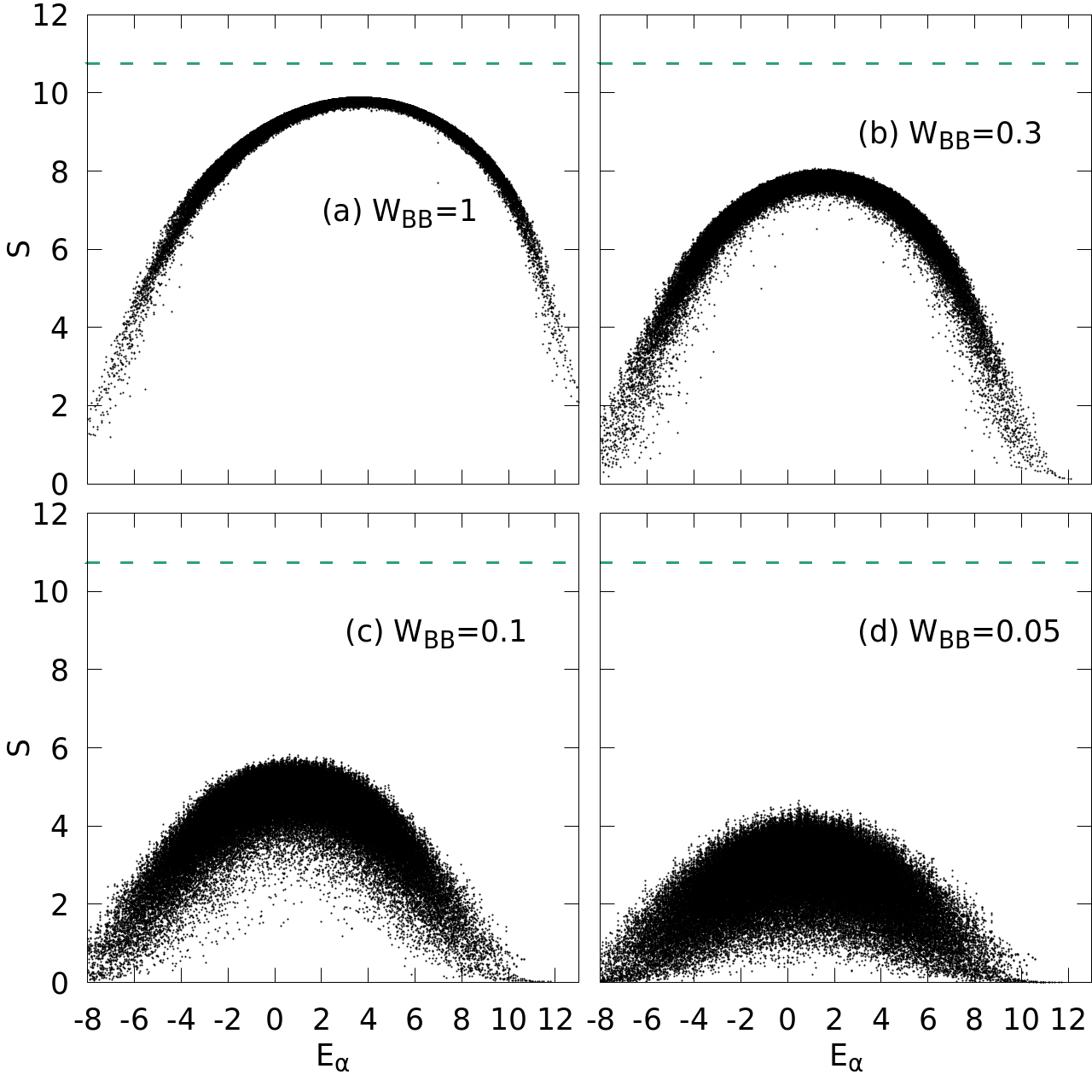}
	\caption{Distribution of Shannon entropies (Eq.~\ref{eq:entropy}) as a measure of the complexity for eigenstates of the system for different $W_{\rm BB}$, (a) $W_{\rm BB}=1$, (b) $W_{\rm BB}=0.3$, (c) $W_{\rm BB}=0.1$, and (d) $W_{\rm BB}=0.05$. The horizontal line marks the value $S_{\rm GOE}\approx\ln{0.48d_{\rm H} }$ expected for the GOE ensemble. All other parameters as in Fig.~\ref{fig:level_stat}.}
	\label{fig:wf_measure}
\end{figure}
As an alternative to spectral measures, on can also explore and quantify chaos through the complexity of the eigenstates itself. According to Berry's conjecture, the eigenstates of a chaotic system feature randomly distributed amplitudes over an appropriate basis, e.g.~in quantum billiards they correspond to randomly distributed plane waves \cite{berry_regular_1977}. Following this conjecture, a large number of such measures have been proposed. They include the statistical distribution of eigenvectors \cite{haake_2001,zelevinsky_nuclear_1996,izrailev_simple_1990, borgonovi_chaos_1998}, the configuration-space probability distribution \cite{mcdonald_wave_1988}, the configuration-space self avoiding path correlation function \cite{shapiro_onset_1984}, the Wigner function based wavefunction autocorrelation function \cite{berry_regular_1977}, the inverse participation ratio \cite{bell_atomic_1970}, the Shannon entropy \cite{blumel_suppression_1984}, and the phase space localization measured in terms of the information entropy encoded in the Husimi distribution \cite{wang_statistical_2020}. One limitation for the quantitative significance of most of these measures (with the possible exception of \cite{wang_statistical_2020}), is their dependence on the chosen basis of representation. For systems that can be continuously tuned from integrable to chaotic, the eigenstates of the integrable limit suggest themselves as a convenient basis to monitor the transition to chaos \cite{zelevinsky_nuclear_1996, santos_onset_2010, borgonovi_chaos_1998}. For many-body systems, the eigenstates of the mean-field Hamiltonian often provide the reference basis for measuring quantum chaoticity \cite{borgonovi_quantum_2016}. In the following, we use the eigenstates $|\psi_\alpha^0\rangle$ of the integrable system with $W_{\rm BB}=0$ as a basis for determining the statistical distribution of eigenvectors. From the amplitudes $c_{\alpha} ^{\alpha'} = \langle \psi_{\alpha'}^0|\psi_\alpha\rangle$ and probabilities $|c_\alpha^{\alpha'}|^2$ we calculate the Shannon entropy \cite{blumel_suppression_1984} for each eigenstate $|\psi_\alpha\rangle$
\begin{equation}\label{eq:entropy}
	S_\alpha = -\sum_{\alpha'=1}^{d_{\rm H}}|c^{\alpha'}_\alpha|^2\ln{|c^{\alpha'}_\alpha|^2}.
\end{equation}
We observe that for $W_{\rm BB}=1$ the Shannon entropy as a function of $E_\alpha$ forms an inverted parabola-like function with remarkably small eigenstate-to-eigenstate fluctuations (Fig.~\ref{fig:wf_measure}). At the apex near the center of the spectrum, $S_\alpha$ reaches a maximum $S_\text{max}$ close to the GOE limit $S_\text{GOE}\approx \ln{0.48d_\text{H}}$ \cite{santos_onset_2010}  with $d_{\rm H}$ the dimension of the Hilbert space [Fig.~\ref{fig:wf_measure} (a)]. States in the tails of the spectrum show strong deviations from this limit as the eigenstates in this region are less complex and do not fulfill the ETH \cite{santos_onset_2010}. Best agreement with GOE predictions can therefore be expected near the center of the spectrum at $\alpha\approx d_{\rm H}/2$ with the highest density of states.\\
For smaller $W_{\rm BB}$ [Fig.~\ref{fig:wf_measure} (b)-(d)] the Shanon entropy reveals a significantly diminished complexity of the eigenstates indicated by a reduced $S_{\rm max}$ and, at the same time, drastically increased state-to-state fluctuations. Probing the generic features of the wavefunctions, we will use the dependence of the scaled Shannon entropy
\begin{equation}\label{eq:barS}
	\bar S = S_{\rm max}(W_{\rm BB})/S_{\rm GOE}
\end{equation}
as an alternative, wavefunction based, measure for the quantum chaoticity complementing the Brody parameter $\gamma$ as spectral measure. Numerically, we determine $S_{\rm max}$ by averaging over small intervals of energy and calculating the maximum of the resulting smooth curve.
\begin{figure}[t]
	\includegraphics[width=\columnwidth]{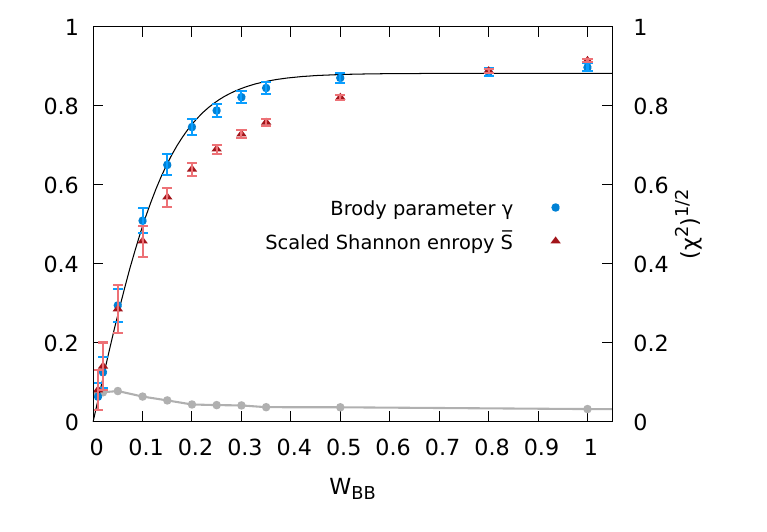}
	\caption{The Brody parameter $\gamma$ (Eq.~\ref{eq:brody}) or scaled Shannon entropy $\bar S$ (Eq.~\ref{eq:barS}) as a function of $W_{\rm BB}$ (left y-axis). The error bars for $\gamma$ correspond to the standard deviation by comparison between the fits to $P(s)$ with fits to $\int_0^s ds'P(s')$, and the black line corresponds to a fit to a $\tanh$ function $\gamma(W_{\rm BB}) \approx \gamma_0\tanh(W_{\rm BB}/W_{\rm BB}^0)$ with the parameters $\gamma_0=0.88$ and $W_{\rm BB}^0=0.15$. The error bars in $\bar S$ reflect the width of $S$ in Fig.~\ref{fig:wf_measure} and correspond to the scaled standard deviation around $S_{\rm max}$. Error of the fit to the Brody distribution as measured by the square root of the $\chi^2$ function (Eq.~\ref{eq:chi2}) (gray line and right y-axis).}
	\label{fig:Brody_entropy}
\end{figure}
Empirically, we find that the dependence of the Brody parameter $\gamma$, i.e.~the degree of quantum chaoticity, on the interaction parameter of the bath particles, $\gamma(W_{\rm BB})$ [Fig.~\ref{fig:Brody_entropy}] can be accurately approximated by
\begin{equation}
	\label{eq:W_BB_ana}
	\gamma(W_{\rm BB}) \approx \gamma_0\tanh(W_{\rm BB}/W_{\rm BB}^0)
\end{equation}
with $\gamma_0=0.88$ and $W_{\rm BB}^0=0.15$. While a monotonic increase is intuitively expected, the origin of this particular simple functional form remains to be understood.
Remarkably, the evolution of $\gamma$ and $\bar S$ as a function of $W_{\rm BB}$ closely mirror each other thereby allowing for two independent measures of the degree of quantum chaoticity during the transition from integrability to chaos. Overall, the agreement between $\gamma$ and $\bar S$ is very good. Residual differences can be viewed as a measure for the residual uncertainty in the quantitative determination of the degree of the eigenstate quantum chaoticity. 
%%%%%%%%%%%%%%%%%%%%%%%%%%%%%%%%%%%%%%%%%%%%%%%%%%%%%%%%%%%%%%%%%%%%%%%%%%
\section{The reduced density matrix of the impurity}\label{sec:1rdm}
%%%%%%%%%%%%%%%%%%%%%%%%%%%%%%%%%%%%%%%%%%%%%%%%%%%%%%%%%%%%%%%%%%%%%%%%%%
The impurity embedded in the Fermi-Hubbard system serves as a   ``thermometer", i.e.~as a sensitive probe of the thermal state of the interacting many-body system. We aim at exploring the emergence of thermal properties of the impurity when the entire (subsystem and bath) system is in a given pure and stationary eigenstate of $\hat H$ with energy $E_\alpha$ and vanishing state entropy (or von Neumann entropy $S_{\rm vN} =0$). Such an isolated large quantum system can be viewed as the limiting case of the quantum microcanonical ensemble where the width of the energy shell $\Delta E$ vanishes, i.e.~$\Delta E\rightarrow 0$. Unlike other approaches, it does not invoke any coarse-graining over a macroscopically small but finite width of the energy shell nor any random interactions. For such a quantum system without any a priori built-in statistical randomness we pose the following question: Starting from a given isolated eigenstate of the entire system, under which conditions will the reduced density matrix of the impurity correspond to a canonical density matrix, i.e.~the thermometer will be accurately represented by a Gibbs ensemble or, for short, be in a Gibbs state? And if such a thermal state emerges, what will be its temperature $T$, or its inverse temperature $\beta=1/k_{\rm B}T$? We refer to this process as emergence of a thermal equilibrium state rather than the frequently used term ``thermalization" as the latter (implicitly) implies a time-dependent approach to an equilibrium state starting from an out-of-equilibrium (statistical or pure) initial state that represents a coherent superposition of different energy eigenstates.  We neither invoke any ensemble average over states from the microcanonical energy shell of finite thickness $\Delta E$ nor do we invoke wave packet dynamics of a non-stationary state of the entire system.\\
For finite isolated systems, in particular, systems with a bounded spectrum such as the present Fermi-Hubbard model, the extraction of proper thermodynamic (or thermostatic) variables from the microcanonical ensemble requires special care. As has been recently demonstrated \cite{dunkel_consistent_2014, hilbert_thermodynamic_2014}, the alternative definitions of the entropy used as the fundamental thermodynamic potential for the microcanonical ensemble yield, in general, inequivalent results. The standard definition \cite{hertz_1910} attributed to Boltzmann
\begin{equation}\label{eq:S_Boltzmann}
	S_{\rm Boltzmann} = k_B\ln{\Omega(E)}=k_{\rm B}\ln{N'(E)},
\end{equation}
with $\Omega(E)$ the DOS of the entire closed system, implies an inverse temperature 
\begin{align}\label{eq:beta_micro_E}
	\beta_{\rm Boltzmann}(E) &= \frac{1}{k_{\rm B}}\frac{\partial S_{\rm Boltzmann}(E) }{\partial E}
	=\frac{\partial\ln{\Omega(E)}}{\partial E}\nonumber \\
	&=\frac{\Omega'(E)}{\Omega(E)} = \frac{N''(E)}{N'(E)},
\end{align}
that may violate certain thermodynamic relations for mesoscopic systems with a bounded spectrum \cite{dunkel_consistent_2014, hilbert_thermodynamic_2014}. As shown more than 100 years ago \cite{Gibbs_1902, hertz_1910}, the Gibbs entropy defined by 
\begin{equation}\label{eq:S_Gibbs}
	S_{\rm Gibbs} = k_B\ln{N(E)}
\end{equation}
results in an inverse temperature 
\begin{equation}\label{eq:beta_Gibbs}
	\beta_{\rm Gibbs}(E) = \frac{\partial\ln{N(E)}}{\partial E} = \frac{N'(E)}{N(E)} = \frac{\Omega(E)}{N(E)}
\end{equation}
that is free of such inconsistencies. From Eq.~\ref{eq:beta_micro_E} and Eq.~\ref{eq:beta_Gibbs} it follows that the two inverse temperature definitions are interrelated through the specific heat $C$ \cite{dunkel_consistent_2014}
\begin{equation}
	\beta_{\rm Boltzmann} = (1-k_B/C)\beta_{\rm Gibbs}
	\label{eq:heat_cap}
\end{equation}
with $C=(\partial T_{\rm Gibbs}/\partial E)^{-1}$ and $T_{\rm Gibbs} = \beta_{\rm Gibbs}^{-1}/k_B$. Only for systems with a small specific heat of the order of $k_{\rm B}$ or smaller, differences between $\beta_{\rm Boltzmann}$
and $\beta_{\rm Gibbs}$ become noticeable. This is in particular the case for systems with a bounded spectrum. While $\beta_{\rm Boltzmann}(E)$ features negative values as soon as the density of states $\Omega(E)=N'(E)$ decreases (Eq.~\ref{eq:beta_micro_E}), $\beta_{\rm Gibbs}(E)$ remains always positive semi-definite (Eq.~\ref{eq:beta_Gibbs}). Fig.~\ref{fig:beta_ana} presents a comparison between $\beta_{\rm Gibbs}$ and $\beta_{\rm Boltzmann}$ for the present Fermi-Hubbard model with an impurity where we have applied the microcanonical thermodynamic relations for $\beta_{\rm Boltzmann}$ and $\beta_{\rm Gibbs}$ (Eqs.~\ref{eq:beta_micro_E}, \ref{eq:beta_Gibbs}) to the numerically determined spectral data (Fig.~\ref{fig:dos}) for the entire system over a wide range of energies $E$. The two inverse temperatures closely follow each other in parallel with $\beta_{\rm Gibbs}$ shifted upwards relative to $\beta_{\rm Boltzmann}$ as long as $\Omega'(E)>0$. For larger $E$ when $\beta_{\rm Boltzmann}$ turns negative, the discrepancies increase as $\beta_{\rm Gibbs}$ remains positive for all $E$.\\
Alternatively, the entire system can be assigned an inverse temperature $\beta_c$ by treating the system as a canonical ensemble. Accordingly, the energy $E$ can be expressed in term of the canonical expectation value
\begin{equation}\label{eq:beta_can}
	E = \frac{{\rm Tr}\left[\hat H e^{-\beta_c \hat H}\right] }{{\rm Tr}\left[ e^{-\beta_c \hat H}\right]}
	= \frac{\partial \ln{Z_c}}{\partial \beta_c},
\end{equation}
where $Z_c={\rm Tr}\left[\exp{(-\beta_c\hat H)}\right]$ is the canonical partition function and $\hat{H}$ is the Hamiltonian of the entire system (see Eq.~\ref{eq:total_ham}). For a given $E$, Eq.~\ref{eq:beta_can} yields an implicit relation for $\beta_c$ also shown in Fig.~\ref{fig:beta_ana}. Obviously, for this finite system $\beta_c$ is close to $\beta_{\rm Boltzmann}$. In the thermodynamic limit we would expect $\beta_c=\beta_{\rm Boltzmann}$. In spite of the fact that the size of our system is still far from the thermodynamic limit $(N\rightarrow \infty)$, the agreement between different thermodynamic ensembles is already remarkably close. Deviations appear primarily near the tails of the density of states and are larger in the region of negative $\beta_{\rm Boltzmann}$ where the DOS decreases rather than increases with $E$.\\
The conceptually interesting question now arises which of these temperatures, if any, will be imprinted on the impurity upon an exact calculation of its reduced density matrix by tracing out all bath degrees of freedom from a given single exact eigenstate of a the isolated many-body system, and without invoking any a priori assumption of the microcanonical ensemble. 
\begin{figure}[t]
	\includegraphics[width=\columnwidth]{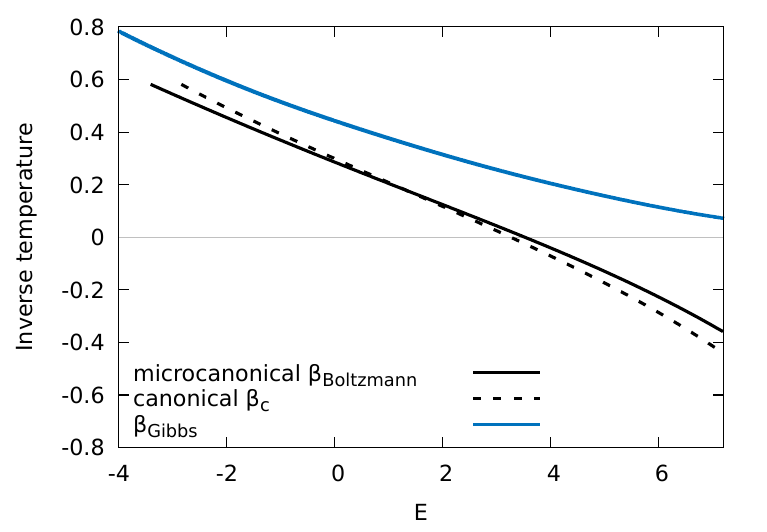}
	\caption{The inverse temperature as a function of the energy of the entire system predicted by the microcanonical Boltzmann entropy (Eq.~\ref{eq:S_Boltzmann}, solid black) and the Gibbs entropy (Eq.~\ref{eq:S_Gibbs}, blue) as well as the canonical expectation value (Eq.~\ref{eq:beta_can}, dashed black). The energy is restricted to the interval $[E_{\rm min},E_{\rm peak}+E_{\rm FWHM}/2,]$ with $E_{\rm min}$ the lower bound where the DOS of the entire system is $\geq15\%$ of its peak value at $E_{\rm peak}$, and $E_{\rm FWHM}$ the full-width-at-half-maximum of the DOS. Bath-bath interaction strength $W_{\rm BB}=1$ and impurity-bath interaction $W_{\rm IB}=1$ [see Fig.~\ref{fig:level_stat} (b)].}
	\label{fig:beta_ana}
\end{figure}
To address this question we start from the density operator for any pure energy eigenstate $|\psi_\alpha\rangle$ of the entire system given by the projector $|\psi_\alpha\rangle\langle \psi_\alpha|$. Consequently, the reduced density matrix (RDM) of the impurity follows from tracing out all bath degrees of freedom,
\begin{eqnarray}\label{eq:1rdm_red}
	D_{\alpha}^{\rm (I)} = {\rm Tr}_{N_{\rm B}}\left[|\psi_\alpha\rangle\langle \psi_\alpha |\right],
\end{eqnarray}
which will, in general, depend on the parent state $|\psi_\alpha\rangle$ it is derived from. We explore now the generic properties of $D_\alpha^{\rm (I)}$ independent of the particular parent state. Specifically, we investigate whether a given $D_\alpha^{\rm (I)}$ emerging from an individual eigenstate $|\psi_\alpha\rangle$ approaches a canonical density matrix. To this end we diagonalize the RDM
\begin{equation}\label{eq:1rdm_diag}
	D_{\alpha}^{\rm (I)} = \sum_{m=1}^{M_s} n_{m,\alpha} |\eta_{m,\alpha}\rangle\langle\eta_{m,\alpha} |,
\end{equation}
yielding natural orbitals $|\eta_{m,\alpha}\rangle$ with natural occupation numbers $n_{m,\alpha}$ \cite{lowdin_natural_1956}. We emphasize that within the present approach the RDMs $D_{\alpha}^{\rm (I)}$ and their eigenvalues, the occupation numbers $n_{m,\alpha}$, which characterize the thermal state, are a priori uniquely determined and not influenced by the choice of any (approximate) basis. Compared to previous investigations, this is one distinguishing feature of the present study of the thermal state emerging from an isolated deterministic many-body system. RDMs have been previously employed in studies of disordered fermionic systems \cite{bera_many-body_2015, bera_oneparticle_2017, lezama_one-particle_2017}.\\ Canonicity is reached when $n_{m,\alpha}$ is given by the Boltzmann factor $e^{-\beta \epsilon_{m,\alpha}^{\rm (I)}}$ with $\epsilon_{m,\alpha}^{\rm (I)}$ the expectation value of the Hamilton operator $H_{\rm I}$ of the impurity alone evaluated in the basis of natural orbitals, $\epsilon_{m,\alpha}^{\rm (I)}=\langle \eta_{m,\alpha}|\hat H_{\rm I}|\eta_{m,\alpha}\rangle$, which, in turn, should be close to the eigenstates of $\hat H_{\rm I}$. Moreover, the resulting value for $\beta$ extracted from the fit to the exponential distribution allows the identification of the inverse temperature uniquely characterizing the thermal distribution.\\
For a finite-size system with an impurity and a bath with an order of magnitude of $10$ particles and finite impurity-bath coupling, the residual interaction of the impurity with the bath is not negligible and should therefore be included to improve the numerical accuracy. We account for the residual impurity-bath interaction on the level of the mean-field (MF) or Hartree approximation \cite{borgonovi_quantum_2016}. Accordingly, the energies $\epsilon^{\rm (I)}$ of the impurity appearing in the Boltzmann factor include a correction term
\begin{equation}\label{eq:nat_orb_ene}
	\bar\epsilon_{m,\alpha}^{\rm (I)}=\langle \eta_{m,\alpha}|\hat H_{\rm I} + \hat W_{\text{MF},\alpha}^{\rm (IB)}|\eta_{m,\alpha}\rangle,
\end{equation}
where the MF interaction operator in site-representation reads
\begin{equation}\label{eq:w_mf}
	 W_{\text{MF},\alpha}^{\text{(IB)}}(j) = W_{\text{IB}}\, \rho_{\text{B},\alpha}(j)
\end{equation}
with 
\begin{equation}\label{eq:rho_b}
	\rho_{\text{B},\alpha}(j)=\langle j| {\text{Tr}}_{N_{\text{B}}-1,{\text{I} }}\left[|\Psi_\alpha\rangle\langle \Psi_\alpha|\right]|j \rangle 
\end{equation}
the reduced one-body density of residual bath particles at the site $j$ when the entire system is in state $|\psi_\alpha\rangle$. In Eq.~\ref{eq:rho_b}, the partial trace over all but one ($N_{\rm B}-1$) bath particles and the impurity ($\text{I}$) is denoted by $\text{Tr}_{N_{\rm B}-1,{\rm I}}$. The energy fluctuations 
\begin{equation}\label{eq:var_e}
	\Delta \bar\epsilon_{m,\alpha}^{\rm (I)} = \sqrt{ \langle \eta_{m,\alpha}|(\hat H_{\rm I} + \hat W_{\text{MF},\alpha}^{\rm (IB)})^2|\eta_{m,\alpha}\rangle - \bar\epsilon_{m,\alpha}^{\rm (I)2} }
\end{equation}
provide a measure for the proximity of the natural orbitals of the RDM to the eigenstates of the (perturbed) single-particle Hamilton operator of the subsystem,  $\hat H_\text{I, eff} = \hat H_{\rm I} + \hat W_{\text{MF},\alpha}^{\text{(IB)}}$. The energy fluctuations (Eq.~\ref{eq:var_e}) vanish only when the natural orbitals $|\eta_{m,\alpha}\rangle$ with which the matrix elements in Eq.~\ref{eq:var_e} are evaluated do coincide with the eigenstate of $\hat{H}_\text{I,eff}$. Therefore, the variance $\Delta \bar\epsilon_{m,\alpha}^\text{(I)}$ can serve a distance measure of the natural orbitals from eigenstates of the impurity Hamiltonian operator. The MF correction in Eq.~\ref{eq:w_mf} follows from the Liouville-von Neumann equation for the reduced system where the interaction with the bath consists of the MF term and a collision operator. The collision operator describes the correlations between the impurity and the bath particles and contains the so-called two-particle (subsystem-bath) cumulant $\Delta_{12}$. We numerically monitor the validity of the MF approximation through the magnitude of the two-particle correlation energy determined by $\Delta_{12}$. Consistently we find that for all many-particle states $|\psi_{\alpha}\rangle$ which reduce to a near-canonical RDM for the impurity, the correlation energy is negligible compared to the MF energy 
thereby justifying Eq.~\ref{eq:nat_orb_ene}. Of course, in the limit of weak impurity-bath coupling, the MF correction (Eq.~\ref{eq:w_mf}) becomes negligible as well. \\
A representative example for the spectrum of the impurity RDM, i.e.~the occupation number distribution of natural orbitals of the impurity RDM emerging from a single energy eigenstate of the entire system with state index $\alpha=4364$ (with $\alpha$ sorted by energy) and energy eigenvalue $E_\alpha = -2.396$ lying on the tail of the DOS with positive $\beta$ for $W_{\rm BB}=1$, is shown in Fig.~\ref{fig:D1_fit}.
\begin{figure}[t]
	\includegraphics[width=\columnwidth]{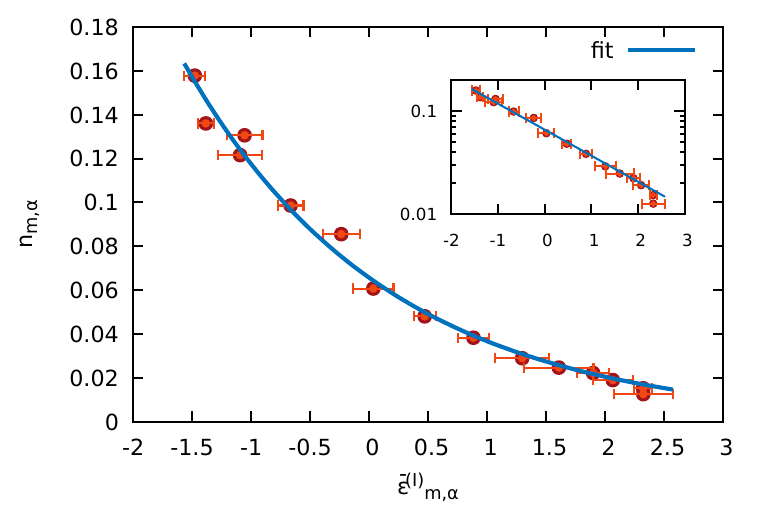}
	\caption{The occupation numbers $n_{m,\alpha}$ of the natural orbitals as a function of their energies $\bar\epsilon_{m,\alpha}$ (Eq.~\ref{eq:nat_orb_ene}) for the eigenstate number $\alpha=4364$ of the total system with energy $E_\alpha \approx -2.396$ and $W_{\rm BB}=1$ ($M_s=15$, $N_{\rm B}=7$). The impurity-bath coupling strength is $W_{\rm IB}=1$. The horizontal error bars indicate the fluctuations $\Delta \bar\epsilon_{m,\alpha}$ (Eq.~\ref{eq:var_e}). The blue solid line corresponds to the best exponential fit yielding the exponent $\beta \approx 0.58$ in agreement with $\beta_{\rm Boltzmann}$ deduced for this state from Eq.~\ref{eq:beta_micro_E}. The inset shows the same plot on a logarithmic scale.}
	\label{fig:D1_fit}
\end{figure}
Indeed, a Boltzmann distribution $\propto e^{-\beta \bar\epsilon_{m,\alpha}^{\rm (I)}}$ characterizing the canonical density matrix is observed. Moreover, the fit to an exponential yields $\beta\approx 0.58$ in close agreement with $\beta_{\rm Boltzmann}=0.58$ predicted by Eq.~\ref{eq:beta_micro_E} for the inverse temperature within the microcanonical ensemble (see also Fig.~\ref{fig:beta_ana}) and reproduces the distribution of occupation numbers very well. It also agrees with $\beta_c$ predicted by Eq.~\ref{eq:beta_can} where the entire system is treated as a canonical ensemble. We note that the Boltzmann-like decay of the diagonal elements would remain qualitatively unchanged when neglecting the MF correction in Eq.~\ref{eq:nat_orb_ene} but the fit to $\beta$ would deteriorate. Thus, from the reduction of state $\alpha=4364$ we have verified that a canonical density matrix emerges.\\
On a conceptual level, the present results confirm the analysis by Dunkel and Hilbert \cite{dunkel_consistent_2014} who showed that the recently observed experimental single-particle population distribution in an isolated finite cold-atom system \cite{braun_negative_2013} is governed by $\beta_{\rm Boltzmann}$. Thus, the canonical density matrix of a small system emerging from tracing out bath variables is characterized by the inverse Boltzmann temperature $\beta_{\rm Boltzmann}$ rather than by $\beta_{\rm Gibbs}$. Consequently, level inversion in a small system in thermal contact with a bath, in particular, spin systems \cite{purcell_1951, ramsay_1965}, can be properly characterized by negative $\beta_{\rm Boltzmann}$. The point to be noted is that while $\beta_{\rm Boltzmann}$ describes the canonical density matrix, the use of $\beta_{\rm Gibbs}$ is required for consistency in thermodynamic relations such as the Carnot efficiency \cite{dunkel_consistent_2014, hilbert_thermodynamic_2014}. In the following, we will present the numerical results for the canonical density matrix of the impurity in therms of $\beta_{\rm Boltzmann}$ which we denote, from now on, for notational simplicity by $\beta$. We point out that $\beta$ can be straightforwardly transformed into $\beta_{\rm Gibbs}$ using Eq.~\ref{eq:heat_cap} and that none of the conclusions to be drawn in the following are altered by this transformation.
%%%%%%%%%%%%%%%%%%%%%%%%%%%%%%%%%%%%%%%%%%%%%%%%%%%%%%%%%%%%%%%%%%%%%%%%%%
\section{Eigenstate canonicity and degree of quantum chaoticity}\label{sec:can_typ}
%%%%%%%%%%%%%%%%%%%%%%%%%%%%%%%%%%%%%%%%%%%%%%%%%%%%%%%%%%%%%%%%%%%%%%%%%%
%
\begin{figure*}[t]
	\includegraphics[width=14.5cm]{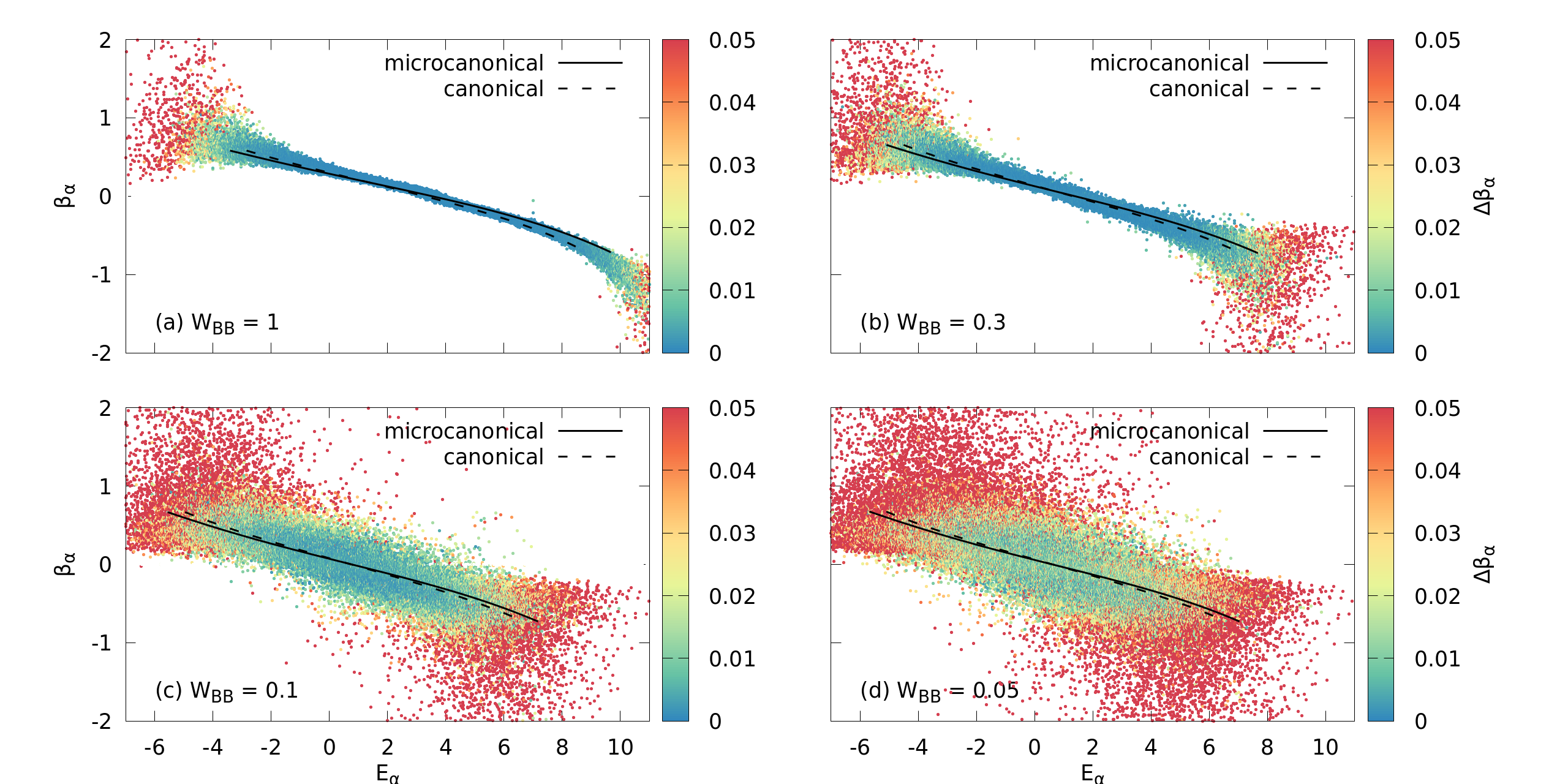}
	\caption{The inverse Boltzmann temperature $\beta_\alpha$ of the impurity as a function of energy $E_\alpha$ for the eigenstates of the entire system as obtained from fits to the RDM of the impurity for varying interaction strengths $W_{\rm BB}$, (a) $W_{\rm BB} = 1$ with $\gamma \approx0.9$, (b) $W_{\rm BB}=0.3$ with $\gamma\approx 0.8$, (c) $W_{\rm BB}=0.1$ with $\gamma\approx 0.5$ and (d) $W_{\rm BB} = 0.05$ with $\gamma\approx0.3$. The color bar on the right hand side represents the variance of $\beta_\alpha$, $\Delta \beta_\alpha$ obtained from the fit, variances above $\Delta \beta > 0.05$ are shown in red. The lines correspond to $\beta(E)$ obtained from the microcanonical ensemble Eq.~\ref{eq:beta_micro_E} (solid) and the canonical ensemble Eq.~\ref{eq:beta_can} (dashed), respectively. Other parameters are $M_s=15$, $N_{\rm B}=7$, $W_{\rm IB}=1$.}
	\label{fig:beta_energy}
\end{figure*}
This demonstration of the emergence of a canonical density matrix from a particular eigenstate $|\psi_\alpha\rangle$ ($\alpha = 4364$) of the entire system invites now the following questions: Is the reduction to a canonical density matrix generic, i.e.~will it emerge for almost all $|\psi_\alpha\rangle$? Is this appearance related to the quantum chaos present in the underlying many-body system? On a more quantitative footing: For how many of the eigenstates will a canonical density matrix emerge and does this number depend on the degree of quantum chaos of the system? \\
We explore these questions by investigating the fraction of many-body eigenstates reducing to a canonical density matrix of the impurity referred to in the following as eigenstate canonicity as a function of the exact total energy $E_\alpha$ for the complete set of eigenstates $\alpha$ of the entire system and for varying bath-bath interaction $W_{\rm BB}$. The corresponding degree of quantum chaoticity of the entire system is measured by either the Brody parameter (Eq.~\ref{eq:brody}) or the Shannon entropy (Eq.~\ref{eq:barS}). Striking differences in the approach to the thermal state with inverse temperature $\beta$ appear which are controlled by the Brody parameter $\gamma$ (or Shannon entropy $\bar S$): At $W_{\rm BB}=1$ when the system is chaotic as indicated by a Brody parameter $\gamma\approx 0.9$ (or scaled Shannon entropy $\bar S=0.9$) a thermal distribution with a well-defined inverse temperature $\beta$, consistent with the (micro)canonical ensemble prediction (Eq.~\ref{eq:beta_micro_E} and Eq.~\ref{eq:beta_can}), emerges for an overwhelming fraction of states with the exception of states in the tails of the spectrum where the DOS is strongly suppressed [Fig.~\ref{fig:beta_energy} (a)]. The large deviations in the tails are consistent with the corresponding deviations for $\bar S$ in the same spectral region [Fig.~\ref{fig:wf_measure} (a)]. With decreasing $W_{\rm BB}$ and, correspondingly, decreasing $\gamma$ or $\bar S$ an increasing fraction of states yield values of $\beta$ that are far from the thermal ensemble prediction. Moreover, the quality of the fit to a canonical density matrix measured by the variance of $\Delta \beta$ and indicated by the color coding of Fig.~\ref{fig:beta_energy} drastically deteriorates.
In other words: for a significant fraction of states, the emerging RDMs do not conform with the constraints of a canonical density matrix.\\
In order to quantify the decomposition of the Hilbert space into the subspace of states $|\psi_\alpha\rangle$ whose reduction to the subsystem yields a canonical density matrix and into the complement whose reduction fails to yield such a thermal state, we introduce a threshold for the variance of the inverse temperature $\Delta \beta_{\rm th}$ above which we consider the eigenstate canonicity to be failing. We then calculate for all states $|\psi_\alpha\rangle$ the fraction of emerging canonical density matrices satisfying $\Delta \beta \leq \Delta \beta_{\rm th}$. Of course, the resulting fraction of states will depend on the precise value of $\Delta \beta_{\rm th}$ chosen. We have determined these fractions for thresholds ranging from $\Delta \beta_{\rm th}=5\times 10^{-3}$ to $1.5\times 10^{-2}$. Changes of the fractions due to variation of $\Delta \beta_{\rm th}$ are indicated by the vertical error bars in Fig.~\ref{fig:cfit_brody}. An unambiguous trend of a monotonic increase of the fraction of canonical density matrices with chaoticity is emerging, obviously unaffected by the choice of $\Delta \beta_{\rm th}$.
\begin{figure}[t]
	\includegraphics[width=\columnwidth]{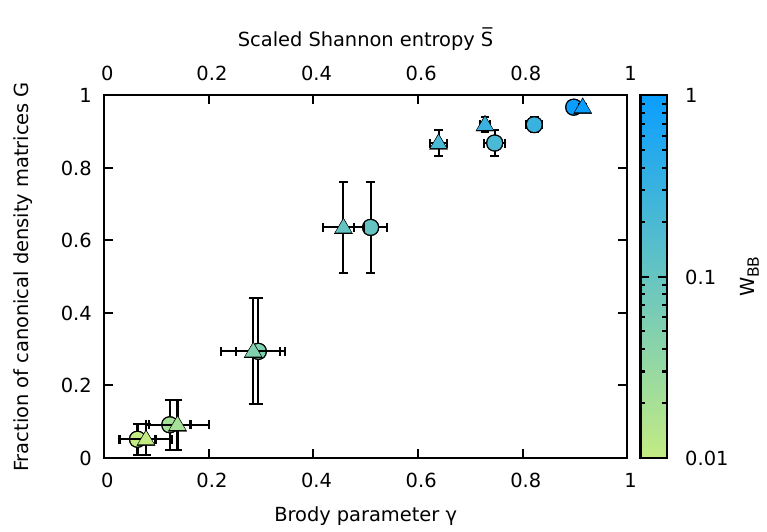}
	\caption{The fraction of canonical density matrices $G$ obtained for the RDM of the impurity as a function of the Brody parameter $\gamma$ (lower horizontal axis, dots) or as a function of the scaled Shannon entropy $\bar S$ (upper horizontal axis, triangles). The dots are color-coded by the interaction strength $W_{\rm BB}$ between the bath particles. Horizontal error bars for $\gamma$ indicate the uncertainty in the extraction of the Brody parameter, horizontal error bars in $\bar S$ indicate the standard deviation of the Shannon entropy (see Fig.~\ref{fig:Brody_entropy}). The vertical error bars give the variation of $G$ under variation of the threshold $\Delta \beta$. Other parameters are $M_s=15$, $N_{\rm B}=7$, and $W_{\rm IB}=1$.}
	\label{fig:cfit_brody}
\end{figure}
This fraction representing Gibbs states, denoted in the following by $G$, monotonically increases with quantum chaoticity as parametrized by either the Brody parameter $\gamma$, $G(\gamma)$, or alternatively by the scaled Shannon entropy, $G(\bar S)$ (Fig.~\ref{fig:cfit_brody}). Since $\gamma$ and $\bar S$ both increase monotonically with the bath interaction $W_{\rm BB}$ (see Fig.~\ref{fig:Brody_entropy}), this implies also a monotonic relationship $G(W_{\rm BB})$. The conceptually important observation emerging from Fig.~\ref{fig:cfit_brody} is that the degree of canonicity of the RDM, $G(\gamma)$, undergoes a continuous transition from the quantum-integrable ($\gamma\rightarrow 0$) to the quantum-chaotic limit ($\gamma \rightarrow 1$). The strength of level repulsion in the NNLS parametrized by $\gamma$ directly determines the probability of finding the RDM of the impurity represented by a Gibbs ensemble.\\
The approach of the RDM of the impurity to the Gibbs ensemble
\begin{equation}\label{eq:gibbs}
	D_{\alpha }^{\rm Gibbs} = \frac{1}{Z_{c,\alpha}}e^{-\beta_\alpha \left(\hat H_I + \hat W_{\text{MF},\alpha}^{\rm (IB)}\right)},
\end{equation}
with $Z_{c,\alpha} = {\rm Tr} [e^{-\beta_\alpha \left(\hat H_I + W_{\text{MF},\alpha}^{\rm (IB)}\right)}]$ can be also directly observed in the spatial site representation $(j_1,j_2)$ of the RDM of the impurity [Fig.~\ref{fig:rdm_sites} (b)].
\begin{figure}[t]
	\includegraphics[width=\columnwidth]{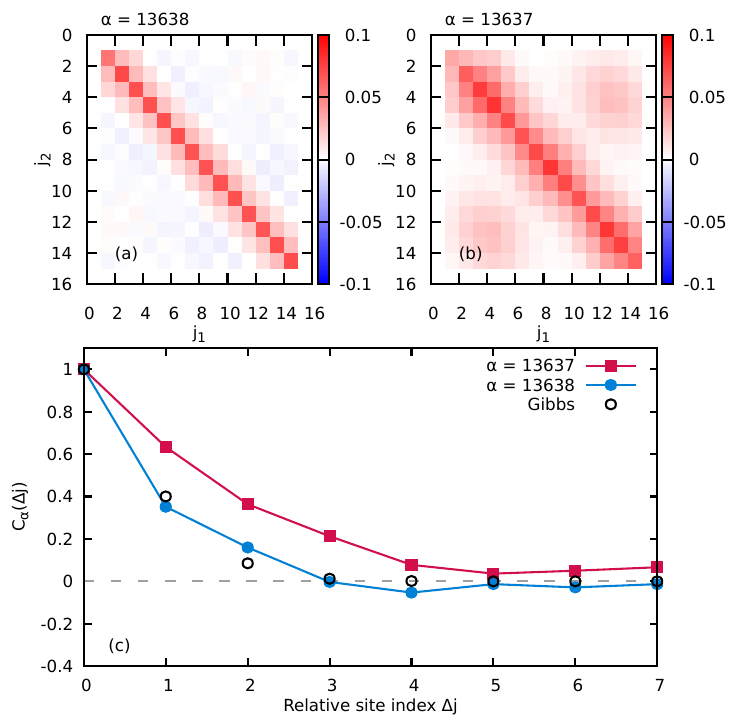}
	\caption{Site representation of the impurity RDM, $\langle j_2|D_\alpha^{\rm (I)}|j_1\rangle$, resulting from the reduction of two adjacent states (a) $\alpha=13638$ and (b) $\alpha=13637$ of the entire system. In (c) we compare the site correlation function $C_\alpha(\Delta j)$ for these two states with the prediction for the ideal Gibbs state (Eq.~\ref{eq:gibbs}) (black open circles). The system is in the transition regime between quantum integrable and quantum chaotic ($W_{\rm BB}=0.1$). Other parameters are $M_s=15$, $N_{\rm B}=7$, $W_{\rm IB}=1$.}
	\label{fig:rdm_sites}
\end{figure}
\begin{figure}[t]
	\includegraphics[width=\columnwidth]{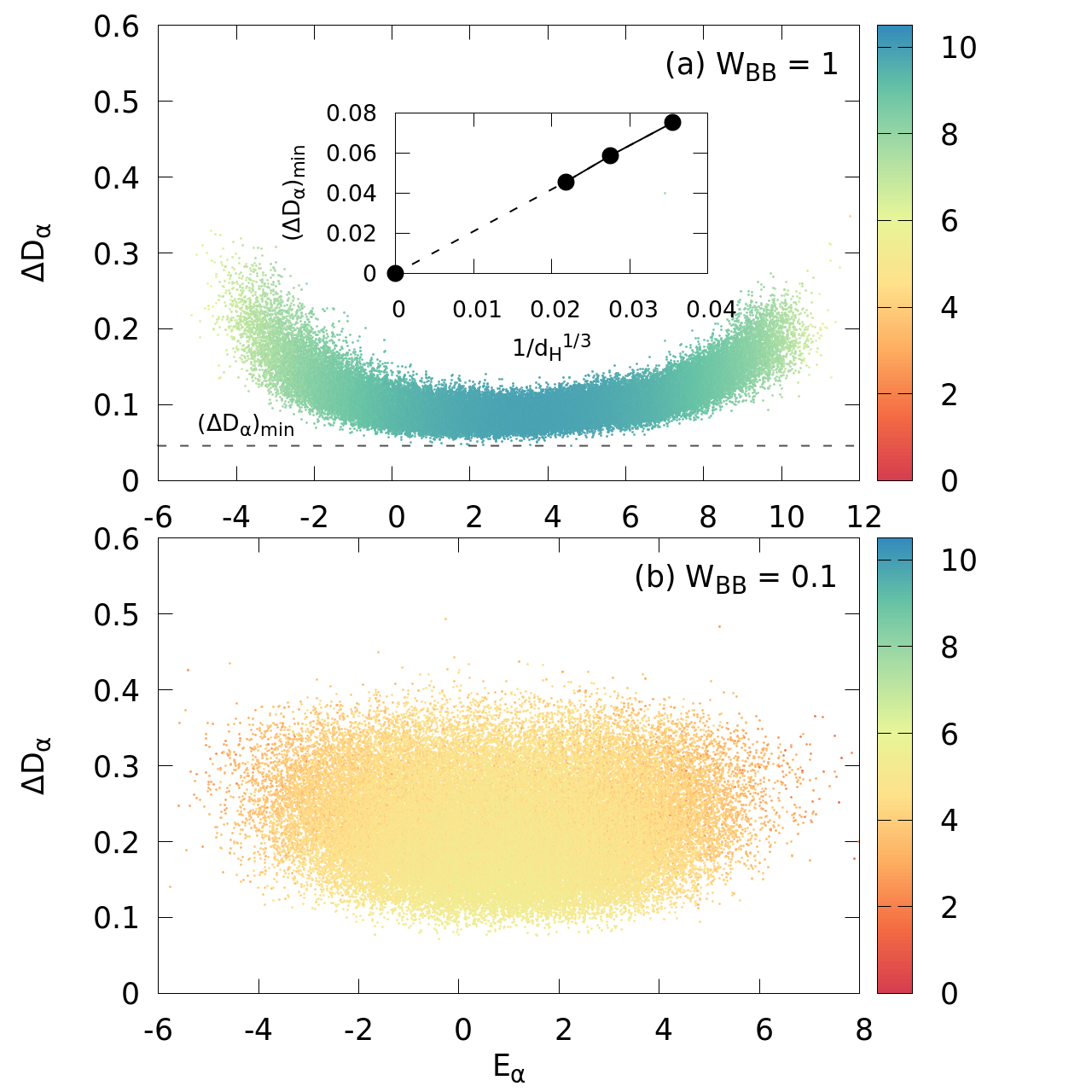}
	\caption{Distribution of distances $\Delta D_\alpha$ (Eq.~\ref{eq:dist_gibbs}) from the (ideal) Gibbs state of the impurity density matrices reduced from the eigenstates $|\psi_{\alpha}\rangle$ of the large system with energy $E_\alpha$. Shown are only those states with variance $\Delta \beta_\alpha<0.01$. The points are color coded by the Shannon entropy of their parent state $|\psi_{\alpha}\rangle$. (a) Near the quantum chaotic limit ($W_{\rm BB}=1$); (b) in the transition regime between quantum integrability and quantum chaos ($W_{\rm BB}=0.1$). The inset in (a) shows the scaling of the minimal distance $(\Delta D_{\alpha})_{\rm min}$ with the dimension of the Hilbert space $d_{\rm H}$ for $W_{\rm BB}=1$. Other parameters are $M_s=15$, $N_{\rm B}=7$, $W_{\rm IB}=1$.}
	\label{fig:distances_to_Gibbs}
\end{figure}
We illustrate the RDM in the site representation for two energetically nearest-neighbor states ($\alpha=13637$ and $\alpha=13638$) when the system is in the transition regime between integrable and non-integrable (in the present case $W_{\rm BB}=0.1$). We quantify the approach to $D_\alpha^{\rm Gibbs}$ through the density matrix site correlation function 
\begin{equation}\label{eq:rdm_cor}
	C_\alpha(\Delta j) = \sum_{j=1}^{M_s-\Delta j} \langle j|D_\alpha^{\rm (I)}|j+\Delta j\rangle,
\end{equation}
where $\langle j|D_\alpha^{\rm (I)}|j'\rangle$ is the RDM of the impurity (Eq.~\ref{eq:1rdm_red}) in the site basis.
While the state $\alpha=13638$ results in a nearly diagonal RDM in the site basis [Fig.~\ref{fig:rdm_sites} (a)] with rapidly decaying site correlations closely following the prediction for a Gibbs ensemble (Eq.~\ref{eq:gibbs}), the adjacent state $\alpha=13637$ yields a RDM with significant off-diagonal entries, extended site correlations, and strong deviations from Eq.~\ref{eq:gibbs}. Thus, the emergence of a thermal density matrix in the transition regime between quantum integrability and quantum chaos 
displays strong state-to-state fluctuations and is not a smooth function of the energy $E_\alpha$.\\
As quantitative measure for the distance of a given RDM from the Gibbs ensemble we use the trace-class norm
\begin{equation}\label{eq:dist_gibbs}
	\Delta D_{\alpha} = ||D_{\alpha}^{\rm (I)}-D_{\alpha}^{\rm Gibbs} ||_1
\end{equation}
with $||M||_1={\text{Tr} \sqrt{M^\dagger M}}$ the largest of the Schatten p-norms ($p=1$). For hermitian positive-semidefinite matrices of unit trace, the Schatten 1-norm is bounded by $0\leq ||M_1-M_2||_1 \leq 2$. We observe for RDMs derived from all eigenstates of the entire system (Fig.~\ref{fig:distances_to_Gibbs}) an overall reduction of distances $\Delta D_\alpha$ from a canonical density matrix with increasing $W_{\rm BB}$. For $W_{\rm BB}=1$ [Fig.~\ref{fig:distances_to_Gibbs} (a)] in the (near) quantum chaotic limit the vast majority of impurity RDMs have a distance of $\lesssim 0.15$ from an ideal Gibbs ensemble (apart from those reduced from many-body states in the tail regions of the spectrum with low DOS). The distribution of $\Delta D_\alpha$ mirrors the distribution of Shannon entropies (Fig.~\ref{fig:wf_measure}). We note that for the present finite quantum system we find that the distance measured by the Schatten 1-norm has a lower bound of $\Delta D_\alpha\gtrsim0.05$. As the Schatten 1-norm is sensitive to small deviations in both diagonal and off-diagonal elements these deviations are due to residual fluctuations (Eq.~\ref{eq:var_e}, Fig.~\ref{fig:D1_fit}) of the natural orbitals of the impurity which are expected to vanish in the thermodynamic limit $N\rightarrow\infty$. Indeed, plotting the value of the smallest distance $(\Delta D_\alpha)_{\rm min}$ as a function of the dimension of the Hilbert space of the system $d_{\rm H}$ for three numerically feasible system sizes indicates that the minimal distance vanishes in the thermodynamic limit as $d_{\rm H}^{-1/3}$ [inset Fig.~\ref{fig:distances_to_Gibbs} (a)].
With decreasing $W_{\rm BB}$, e.g.~$W_{\rm BB}=0.1$ in Fig.~\ref{fig:distances_to_Gibbs} (b), the mean distance of RDMs from a Gibbs state significantly increases and, moreover, the spread becomes much larger reflecting, again, the behavior of the Shannon entropy [Fig.~\ref{fig:wf_measure} (c)]. \\
The emergence of canonical density matrices, i.e.~of Gibbs states for almost all $|\psi_\alpha\rangle$ in the quantum chaotic limit ($\gamma\simeq 1$ or $\bar S\simeq1$) can be viewed as a rather specific realization of the ETH \cite{deutsch_quantum_1991, srednicki_approach_1999, srednicki_chaos_1994, srednicki_thermal_1996}. The local observable in this case is the RDM of the impurity, $D^{(\text{I})}_\alpha$, itself. Its diagonal elements are a smooth function of the total energy $E_\alpha$ as predicted by ETH but now, more specifically, Boltzmann distributed $\propto e^{-\beta_\alpha \bar\epsilon_{m,\alpha}^{\rm (I)}}$ over impurity states with the inverse temperature imprinted by $E_\alpha$. The present analysis covers, in addition, also the transition regime between quantum integrability and quantum chaos ($0<\gamma<1$) where, in general, the ETH does not apply. A canonical density matrix may still emerge but now only for a decreasing fraction of eigenstates of the finite large system. The size of this fraction $G$ is predicted by the degree of quantum chaoticity as measured by the Brody parameter $\gamma$ of Shannon entropy $\bar S$ (Fig.~\ref{fig:cfit_brody}).\\
The direct relation between the emergence of the canonical density matrix for a small subsystem from eigenstate reduction and the quantum chaoticity of the large system it is embedded in, established here for a finite quantum system, raises the conceptual question as to the extension of this connection to the thermodynamic ($N\rightarrow\infty$) limit. Clearly, this question cannot be conclusively addressed by the present method of exact diagonalization. Nevertheless, we can provide evidence to this effect by exploring the scaling with system size still within computational reach.
We first establish that the degree of quantum chaoticity as measured by the Brody parameter $\gamma$ (or the Shannon entropy $\bar S$), indeed, increases with system size at fixed strength of the interaction $W_{\rm BB}$ that breaks quantum integrability (Fig.~\ref{fig:sys_size}).
\begin{figure}[t]
	\includegraphics[width=\columnwidth]{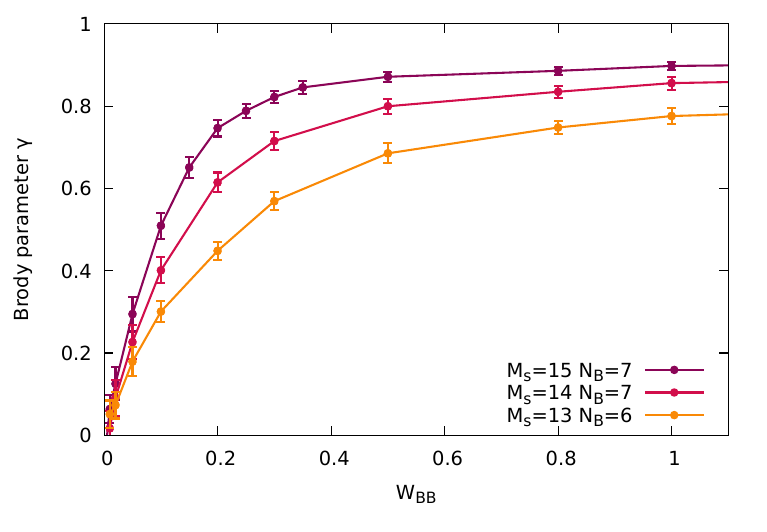}
	\caption{Variation of the Brody parameter $\gamma$ characterizing the transition from quantum integrability to quantum chaos as a function of the bath-bath interaction $W_{\rm BB}$ breaking quantum integrability shown for different system sizes. The impurity-bath interaction is $W_{\rm IB}=1$.}
	\label{fig:sys_size}
\end{figure}
We vary the system size by increasing the total number of sites while keeping the system at (approximate) half-filling of bath particles. The corresponding Hilbert space increases from $d_{\rm H}=22308$ ($M_s=13$, $N_{\rm B}=6$) to $d_{\rm H}=96525$ ($M_s=15$, $N_{\rm B}=7$). The observed increase of quantum chaoticity with system size is qualitatively in line with properties of classical chaos: In a mixed phase space with surviving local regular structures such as tori, their influence on phase space dynamics is rapidly diminishing with increasing phase space dimension a prominent example of which is Arnold diffusion \cite{arnold_1979, lichtenberg_1991}. This increase of quantum chaoticity with system size at fixed interaction strength turns out to be key for the emergence of a universal, i.e.~(nearly) system-size independent, interrelation between the fraction of canonical eigenstates and the degree of quantum chaoticity. Both the Brody parameter $\gamma$ as well as the fraction of density matrices complying with the Gibbs ensemble increase with system size at fixed bath interaction strength. As a consequence, a near universal, i.e.~size independent, relation $G(\gamma)$ between the fraction of (approximate) Gibbs states and the degree of quantum chaos as measured by $\gamma$ emerges (Fig.~\ref{fig:sigma_brody_sys}). 
\begin{figure}[t]
	\includegraphics[width=\columnwidth]{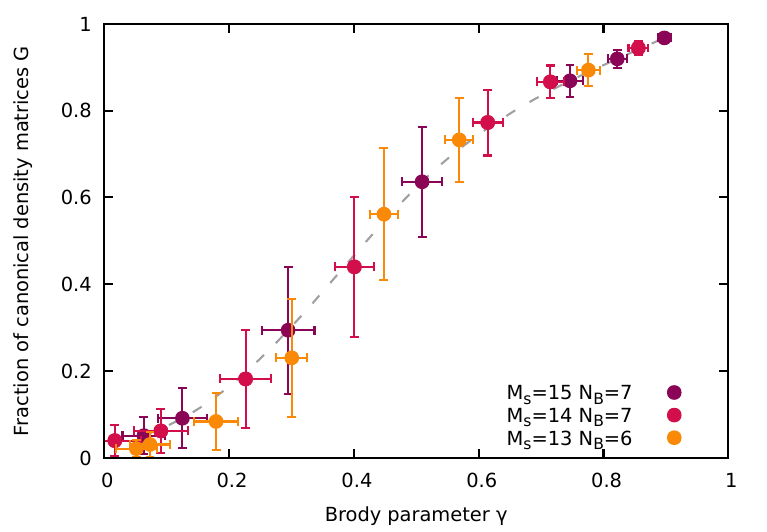}
	\caption{Universal relation between the fraction $G$ of RDMs of the impurity converging to a Gibbs state and the Brody parameter $\gamma$ for different combinations of system sizes ($M_s$ and $N_{\rm B}$) and bath interaction strengths $W_{\rm BB}$. Impurity-bath interaction in all systems considered is $W_{\rm IB}=1$. Dashed line to guide the eye.}
	\label{fig:sigma_brody_sys}
\end{figure}
The data for different combinations of values of $W_{\rm BB}$ and $M_s$ fall on the same curve. A very similar relation would emerge for $G(\bar S)$ as a function of the scaled Shannon entropy. We have thus established the remarkable feature that the fraction of canonical eigenstates, i.e.~the likelihood that a subsystem is in a Gibbs state when the large but finite system is in a pure energy eigenstate with zero von Neumann entropy is controlled and can be tuned by $\gamma$ (or $\bar S$) and, in turn, by the degree of level repulsion in the quantum many-body system which is controlled by $\gamma$. 
%%%%%%%%%%%%%%%%%%%%%%%%%%%%%%%%%%%%%%%%%%%%%%%%%%%%%%%%%%%%%%%%%%%%%%%%%%
\section{Conclusions and outlook}\label{sec:conc}
%%%%%%%%%%%%%%%%%%%%%%%%%%%%%%%%%%%%%%%%%%%%%%%%%%%%%%%%%%%%%%%%%%%%%%%%%%
In this work we have explored the emergence of a thermal state (or Gibbs ensemble) of a small (sub)system in contact with a bath when the combined large but finite deterministic quantum system is isolated and in a well-defined energy eigenstate. 
As prototypical case we have considered an impurity embedded in an interacting spin-polarized Fermi-Hubbard many-body bath which facilitates a clear-cut subsystem-bath decomposition and a tunable transition of the entire system from quantum integrability to quantum chaos. By tracing out the bath degrees of freedom, we have investigated how many of the resulting reduced density matrices of the subsystem represent a canonical density matrix. We have shown that the probability for finding a canonical density matrix monotonically increases with the degree of quantum chaos. The degree of quantum chaos is identified here by both the energy-level statistics as well as by the randomness of the eigenstates as measured by the Shannon entropy. The likelihood for the emergence of thermal states is thus found to be controlled by the degree of quantum chaoticity as parametrized by the Brody parameter or the Shannon entropy. Even though our simulations are limited to finite-size systems, the present results for varying system sizes suggest that the relation between the fraction of eigenstates of the isolated many-body system whose reduction to a small subsystem yields a reduced canonical density matrix and the degree of quantum chaoticity is universal, i.e.~size-independent.\\
Each many-body eigenstate represents the fine-grained version of the energy shell of the microcanonical ensemble of the entire impurity-bath system. This connection between the fraction of canonical eigenstates and quantum chaoticity thus offers a direct quantum analogue to the role of classical chaos which Boltzmann invoked in deducing the classical (micro-)canonical ensemble.
One can view this as an example of classical-quantum correspondence to this cornerstone of the foundation of statistical mechanics.
The statistical ensemble properties can already emerge for isolated energy eigenstates without invoking any randomness, e.g.~coarse-graining over a macroscopically thin energy shell or superposition of many eigenstates of the isolated large system as frequently employed.
The emergence of statistical ensemble properties from the reduction of pure states was already early anticipated by Landau and Lifshitz \cite{landau_1958} and later related to quantum chaos \cite{borgonovi_quantum_2016}. The present study establishes a direct quantitative relationship between the degree of canonicity and the degree of quantum chaos in particular also covering the transition regime from quantum integrability to quantum chaos.
\\
The present results are also expected to have implications for the topical issue of thermalization in finite quantum systems \cite{jansen_eigenstate_2019, neill_ergodic_2016, prufer_observation_2018, erne_universal_2018}. In this paper, we intentionally avoided this notion and, instead, focused on thermal equilibrium states as we deduce the canonical density matrix from stationary energy eigenstates bypassing any explicit time dependence of the dynamics. Thermalization of an initial non-equilibrium state is, by contrast, a fundamental probe of the time evolution of quantum many-body systems. Up to now, one primary focus has been on quantum quenches, the relaxation of out-off equilibrium initial states. Their time evolution has, typically, shown a transition from an exponential decay for weakly perturbed many-body systems to a Gaussian decay in the strongly coupled limit, however, without an unambiguous correlation to quantum chaos \cite{torres-herrera_local_2014, torres-herrera_relaxation_2015}. The Shannon entropy was found to linearly increase with time before reaching saturation \cite{santos_onset_2012}.
For disordered systems, an initial rapid decay followed by a slow power-law relaxation of occupation numbers has been observed \cite{lezama_one-particle_2017, krause_nucleation_2021}.
The extension of the present study to a non-equilibrium initial state of a deterministic many-body system would yield the time evolution of the entire one-body RDM, and its eigenvalues and eigenvectors, the time dependence of which remains to be explored. Moreover, the dependence of the relaxation dynamics of the RDM on the choice of the initial state for systems in the transition regime between quantum integrability and quantum chaos (i.e.~for intermediate values of the Brody parameter $\gamma$) is of particular interest. Most importantly, will quantum chaos play an analogous role for the process of mixing as classical chaos does for classical non-equilibrium dynamics and the relaxation to equilibrium? The origin and properties of such ``quantum mixing" remain a widely open question.

%%%%%%%%%%%%%%%%%%%%%%%%%%%%%%%%%%%%%%%%%%%%%%%%%%%%%%%%%%%%%%
%\authorcontributions{Conceptualization, I.B., S.D. and F.L.; methodology, I.B. and F.L.; software, I.B., M.K., S.D. and F.L.; validation, I.B. and J.B.; formal analysis, I.B. and J.B.; investigation, I.B.; resources, I.B.; data curation, I.B. and M.K.; writing---original draft preparation, I.B.; writing---review and editing, I.B., M.K., S.D., F.L. and J.B.; visualization, I.B.; supervision, I.B.; project administration, I.B. and J.B.; funding acquisition, I.B. and J.B. All authors have read and agreed to the published version of the manuscript.}

\section*{Acknowledgments}
We thank Sebastian Bichelmaier for helpful discussions.This research was funded by the WWTF grant MA-14002, the Austrian Science Fund (FWF) grant P 35539-N, the FWF doctoral college Solids4Fun, as well as the International Max Planck Research School of Advanced Photon Science (IMPRS-APS). Calculations were performed on the Vienna Scientific Cluster (VSC3 and VSC4).

%%%%%%%%%%%%%%%%%%%%%%%%%%%%%%%%%%%%%%
\section*{References}
%\bibliography{thermalization_biblio}

\end{document}